\UseRawInputEncoding
\documentclass[twocolumn]{aastex62}
\usepackage{afterpage}
\usepackage{longtable}
\usepackage{amsmath}
\usepackage{hyperref}

\accepted{Dec 24, 2021}

\newcommand{\Lya}{\hbox{{\rm Ly}$\alpha$}}
\newcommand{\Ha}{\hbox{{\rm H}$\alpha$}}
\newcommand{\Hb}{\hbox{{\rm H}$\beta$}}

\newcommand{\HeII}{\hbox{{\rm He}\kern 0.1em{\sc ii}}}
\newcommand{\OII}{\hbox{{\rm [O}\kern 0.1em{\sc ii}{\rm ]}}}
\newcommand{\OIII}{\hbox{{\rm [O}\kern 0.1em{\sc iii}{\rm ]}}}
\newcommand{\CIV}{\hbox{{\rm C}\kern 0.1em{\sc iv}}}
\newcommand{\CIII}{\hbox{{\rm C}\kern 0.1em{\sc iii}{\rm ]}}}
\newcommand{\MgII}{\hbox{{\rm Mg}\kern 0.1em{\sc ii}}}
\newcommand{\FeII}{\hbox{{\rm Fe}\kern 0.1em{\sc ii}}}
\newcommand{\Mbh}{\ensuremath{M_{\rm BH}}}

\newcommand{\Mdot}{\ensuremath{{\dot{M}}_{\rm BH}}}
\newcommand{\lum}{\ensuremath{\lambda L_{3000}}}

\newcommand{\cream}{\texttt{CREAM}}
\newcommand{\jav}{\texttt{JAVELIN}}
\newcommand{\pycecream}{\texttt{PyceCREAM}}
\newcommand{\hst}{\textit{HST}}

\begin{document}

\title{The Sloan Digital Sky Survey Reverberation Mapping Project: UV-Optical Accretion Disk Measurements with Hubble Space Telescope}
\correspondingauthor{Yasaman Homayouni}
\email{yhomayouni@stsci.edu}

\author[0000-0002-0957-7151]{Y. Homayouni}
\affil{University of Connecticut, Department of Physics, 2152 Hillside Road, Unit 3046, Storrs, CT 06269-3046, USA}
\affiliation{Space Telescope Science Institute, 3700 San Martin Drive, Baltimore, MD 21218, USA}

\author[0000-0003-1055-1888]{Megan R. Sturm}
\affiliation{University of Connecticut, Department of Physics, 2152 Hillside Road, Unit 3046, Storrs, CT 06269-3046, USA}
\affiliation{Department of Physics, Montana State University, Bozeman, MT 59717, USA}

\author[0000-0002-1410-0470]{Jonathan R. Trump}
\affiliation{University of Connecticut, Department of Physics, 2152 Hillside Road, Unit 3046, Storrs, CT 06269-3046, USA}

\author[0000-0003-1728-0304]{Keith Horne}
\affiliation{SUPA Physics and Astronomy, University of St. Andrews, Fife, KY16 9SS, Scotland, UK}

\author[0000-0001-9920-6057]{C. J. Grier}
\affiliation{Steward Observatory, The University of Arizona, 933 North Cherry Avenue, Tucson, AZ 85721, USA}

\author[0000-0001-8610-5732]{Yue Shen}
\affiliation{Department of Astronomy, University of Illinois at Urbana-Champaign, Urbana, IL, 61801, USA}
\affiliation{National Center for Supercomputing Applications, University of Illinois at Urbana-Champaign, Urbana, IL, 61801, USA}

\author[0000-0002-0167-2453]{W. N. Brandt}
\affiliation{Dept. of Astronomy and Astrophysics, The Pennsylvania State University, 525 Davey Laboratory, University Park, PA 16802}
\affiliation{Institute for Gravitation and the Cosmos, The Pennsylvania State University, University Park, PA 16802}
\affiliation{Department of Physics, 104 Davey Lab, The Pennsylvania State University, University Park, PA 16802, USA}

\author{Gloria Fonseca Alvarez}
\affil{University of Connecticut, Department of Physics, 2152 Hillside Road, Unit 3046, Storrs, CT 06269-3046}

\author[0000-0002-1763-5825]{P. B. Hall}
\affiliation{Department of Physics and Astronomy, York University, Toronto, ON M3J 1P3, Canada}

\author[0000-0001-6947-5846]{Luis C. Ho}
\affiliation{Kavli Institute for Astronomy and Astrophysics, Peking University, Beijing 100871, China}
\affiliation{Department of Astronomy, School of Physics, Peking University, Beijing 100871, China}

\author[0000-0002-0311-2812]{Jennifer I-Hsiu Li}
\affiliation{Department of Astronomy, University of Illinois at Urbana-Champaign, Urbana, IL, 61801, USA}
\affiliation{University of Michigan, Ann Arbor, MI 48109, USA}

\author[0000-0002-0771-2153]{Mouyuan Sun}
\affiliation{Dept. of Astronomy, Xiamen University, Xiamen, Fujian 361005, China}

\author{D. P. Schneider}
\affiliation{Dept. of Astronomy and Astrophysics, The Pennsylvania State University, 525 Davey Laboratory, University Park, PA 16802}
\affiliation{Institute for Gravitation and the Cosmos, The Pennsylvania State University, University Park, PA 16802}

\begin{abstract}
We present accretion-disk structure measurements from UV-optical reverberation mapping observations of a sample of eight quasars at $0.24<z<0.85$. Ultraviolet photometry comes from two cycles of \textit{Hubble Space Telescope} monitoring, accompanied by multi-band optical monitoring by the Las Cumbres Observatory network and Liverpool Telescopes. The targets were selected from the Sloan Digital Sky Survey Reverberation Mapping (SDSS-RM) project sample with reliable black-hole mass measurements from \Hb\ reverberation mapping results. We measure significant lags between the UV and various optical $griz$ bands using \jav\ and \cream\ methods. We use the significant lag results from both methods to fit the accretion-disk structure using a Markov chain Monte Carlo approach. We study the accretion disk as a function of disk normalization, temperature scaling, and efficiency. We find direct evidence for diffuse nebular emission from Balmer and \FeII\ lines over discrete wavelength ranges. We also find that our best-fit disk color profile is broadly consistent with the Shakura\,\&\,Sunyaev disk model. We compare our UV-optical lags to the disk sizes inferred from optical-optical lags of the same quasars and find that our results are consistent with these quasars being drawn from a limited high-lag subset of the broader population. Our results are therefore broadly consistent with models that suggest longer disk lags in a subset of quasars, for example, due to a nonzero size of the ionizing corona and/or magnetic heating contributing to the disk response.
\end{abstract}

\section{Introduction} \label{sec:intro}
Although many advances in observing active galactic nuclei (AGN) have been made, the detailed physics of accretion onto the central engine, the supermassive black hole (SMBH), remains poorly understood. The classic solution for an accretion disk around a compact object is described by \citet{SS1973}. The gas infall around a black hole was modeled by a geometrically thin, optically thick accretion disk (hereafter SS73). An effective viscosity causes gas to spiral inwards and converts some of its potential energy into thermal radiation. If the disk is optically thick, the local thermal emission, at least approximately, corresponds to black body radiation leading to a continuum emission spectrum, which peaks at ultraviolet (UV) wavelengths in a typical AGN spectral energy distribution.

In the ``lamp-post" model, the disk is directly illuminated by an extreme-UV and X-ray ionizing source above/below the disk \citep{Galeev1979, Krolik1991, Reynolds2003}. The ionizing radiation is reprocessed by the disk surface, starting with the inner disk and propagating outward to the outer disk, allowing for coherent continuum variations at different radii \citep{Cackett2007}. The lamp-post reprocessing enables the use of correlated inter-band variability signatures to measure the accretion-disk size and structure from the light travel time (i.e., $\tau$) between the short and long-wavelength emission from the disk. This is the basic assumption of the reverberation mapping (RM) technique \citep{Blandford1982, Peterson1993, Peterson2004} in which physically connected regions ``reverberate" in response to the driving continuum. The RM technique has been widely used to estimate the size of the broad-line emitting region (BLR), and subsequently the SMBH mass from the virial product (\citealt{Bentz2015}, usually known as broad-line RM).

Alternatively, the RM technique can be applied to infer accretion-disk size, commonly known as continuum RM.
The near-UV and optical continuum varies in response to the unobserved far-UV and X-ray ionizing continuum after a time delay. Continuum RM enables studies of accretion-disk structure by measuring the time delay of causally connected regions of the accretion disk. The continuum RM technique has proved to be more challenging compared to broad-line RM. This is largely because accretion disks are smaller than the BLR, so continuum lags are typically much smaller than broad-line lags.
Nevertheless, continuum RM is the most promising technique to learn about SMBH accretion physics for quasars in the distant Universe.

Early continuum RM studies established the stratified temperature profile of accretion disks, showing cooler material at larger radii \citep{Krolik1991,Wanders1997,Collier1998,Collier2001}. Several recent monitoring campaigns have been dedicated to accretion-disk studies using continuum RM in nearby AGNs \citep{Sergeev2005, Shappee2014, McHardy2014, Edelson2015, Fausnaugh2016, Edelson2017, Fausnaugh2018, McHardy2018}. The results indicate a strong correlation of lightcurve variability in the UV-optical with UV variations leading those at optical wavelengths. The general trend in disk-temperature profile (i.e., the wavelength scaling) through the continuum emission from inner/hotter to outer/cooler disk regions is consistent with the lamp-post model \citep{Cackett2007} with $\tau \propto \lambda^{4/3}$ as expected by \citet{SS1973}.

Most accretion-disk sizes measured from continuum RM are significantly larger than the expectation from the SS73 model. Observations of single, local AGN have reported UV-optical lags that are a factor of $\sim$2-3 larger than the model expectation \citep{McHardy2014, Edelson2015, Fausnaugh2016, Edelson2017,McHardy2018}. Similarly, microlensing observations suggest disk sizes that are $\sim3\times$ larger than the SS73 disk-size expectation \citep{Morgan2018}.
However, multi-object continuum RM measurements of higher redshift (up to $z\approx 1.9$) quasars are mixed \citep{JiangGreen2017, Mudd2018, Homayouni2019, Yu2020}.
Larger than expected UV-optical lags have also been reported for interband optical continuum RM lags from PAN-STARRS \citep{JiangGreen2017}. Other works on interband optical continuum RM \citep{Mudd2018, Homayouni2019, Yu2020} challenge this common picture, reporting a consistent accretion-disk size with the \citet{SS1973} model; for example, \citet{Mudd2018, Yu2020} relax the lag-significance criteria to compute the disk size directly from the interband optical lightcurves.

The Sloan Digital Sky Survey Reverberation Mapping Project (SDSS-RM, \citealt{Shen2015a}) has been effective in the industrial-scale study of 849 quasars at $z>0.3$, spanning a diverse quasar population in redshift, mass, and accretion rate \citep{Shen2019a}. Recently, \citet{Homayouni2019} used the SDSS-RM survey and a Markov chain Monte Carlo approach to fit accretion-disk structure and included lag-detection limits to avoid biases in the measured disk sizes; for more discussion, see the Appendix in \citet{Homayouni2019}.

The present work describes the results of an intensive, multiwavelength monitoring campaign for eight quasars selected from the SDSS-RM parent sample. We obtained UV monitoring observations from the \textit{Hubble Space Telescope} (\hst) and coordinated ground-based optical monitoring from the \textit{Liverpool Telescope} and \textit{Las Cumbres Observatory}. This study includes a diverse sample of quasars in terms of black-hole mass (\Mbh) and accretion rate with UV-optical broadband photometric monitoring beyond the local universe $z>0.1$. The present work has two primary goals. The first is to measure the UV emission from the accretion disk's inner regions and compare the differences in the measured disk sizes for a diverse sample of quasars. The second goal is to measure the UV-optical lag to map the stratification of accretion-disk structure.

In Section~\autoref{sec:data} we discuss the details of the observations. Section~\autoref{sec:Reduction} illustrates our custom reduction pipeline. In Section~\autoref{sec:timeseries} we describe our lag-identification method, lag reliability, and alias removal for each individual target. In Section \autoref{sec:discussion} we present our final UV disk size and accretion-disk model fits and connection to mass and accretion rate. Throughout this work, we adopt a $\Lambda$CDM cosmology with $\Omega_{\Lambda}$ = 0.7, $\Omega_M$ = 0.3, and $H_0$ = 70 km $\mathrm{s^{-1} Mpc^{-1}}$.
\section{Observations} \label{sec:data}
\begin{figure}[t]\label{Fig1:sample}
\centering
\includegraphics[width=88mm]{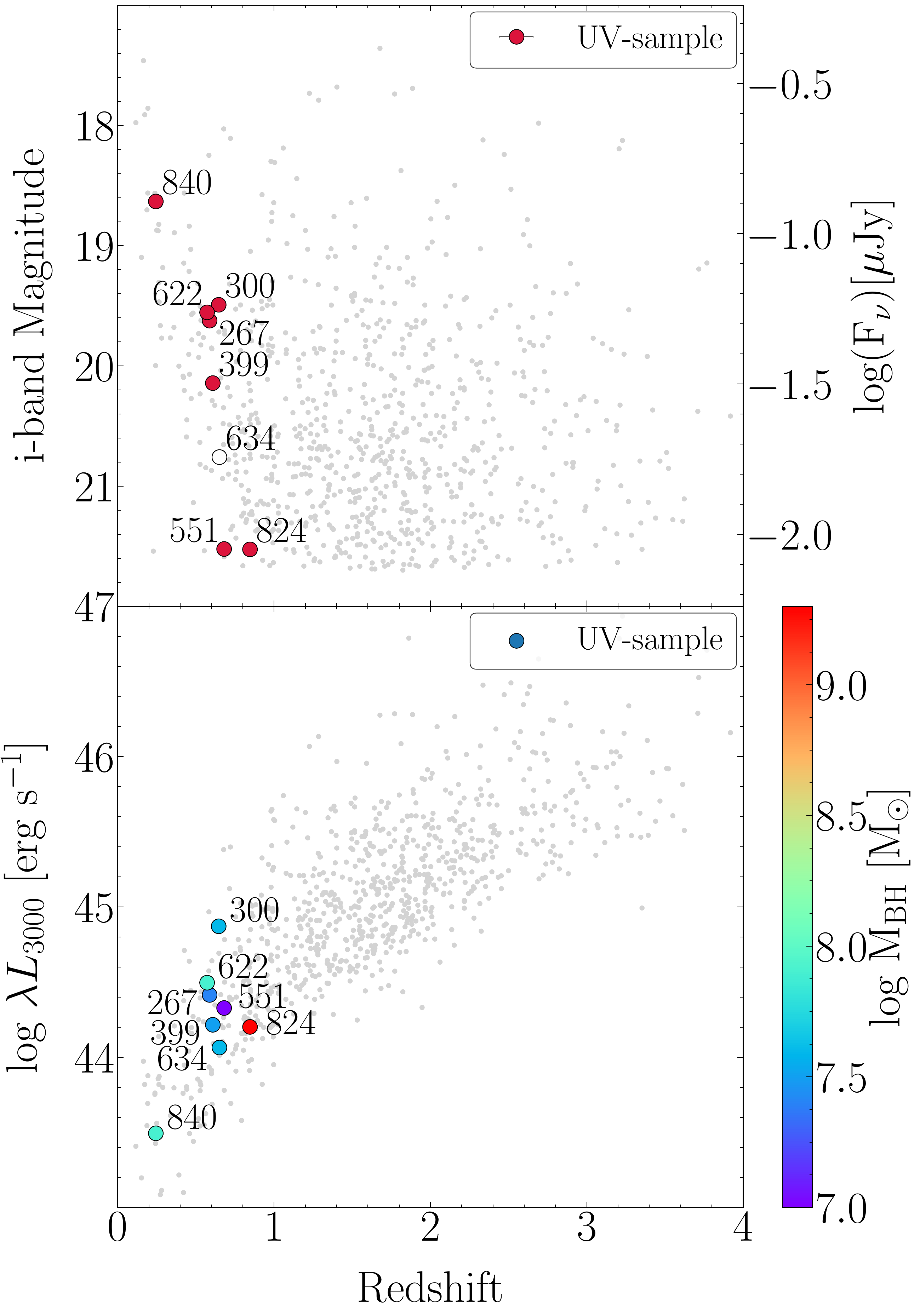}
 \caption{\textbf{Top:} The SDSS-RM parent sample of 849 quasars (gray points) and the set of eight quasars from the UV monitoring campaign (red filled points). The open symbol illustrates RM634, which had poor signal-to-noise as the result of a DASH observing mode (see Section \ref{sec:dash}). \textbf{Bottom:} Our targets probe a wide range of quasar parameter space in \lum\ continuum luminosity and black-hole mass, as established for the broader sample in \citet{Shen2019a}, compared to UV-optical accretion disk studies for local AGNs.
 }
\label{fig1}
\end{figure}

Our set of eight targets for this study is drawn from the 849 quasars of the SDSS-RM sample (see Table~\ref{tab:table1}). These targets are significantly variable with fractional continuum root mean square (RMS) variability of 10\%-50\% measured from the Prepspec software \citep{Shen2015a, Shen2016a} at rest-frame \lum\ continuum. These targets probe a broad range of quasar parameter space in redshift, mass, and Eddington ratio.
All targets in our sample have reverberation mapping \Mbh\ measurements from the \Hb\ emission line \citep{Grier2017}, except RM824. We used the single-epoch mass for this particular target, as reported in \citet{Shen2019a}. Additionally, we selected our targets to have $<10\%$ BLR contamination in the WFC3 F275W filter. This is to minimize the effects of strong emission lines on broadband filters, since broad emission lines typically have longer timescales of variability with longer lags and may bias the continuum lightcurves' underlying shorter lags. Table~\ref{tab:table1} gives a brief description of our selected sample properties. Figure~\ref{Fig1:sample} illustrates the probed quasar parameter space in \textit{i}-mag, redshift, luminosity, and black-hole mass.

\afterpage
\startlongtable
\begin{deluxetable*}{cccccccccc}
\tablecaption{Quasar Sample Information\label{tab:table1}}
\tablehead{
\colhead{RMID} & \colhead{R.A.} & \colhead{Decl.} & \colhead{$z$} & \colhead{$i$mag} & \colhead{Var${}^a$} & \colhead{log $\lambda L_{3000}$} & \colhead{$\log \Mbh^{b}$} & \colhead{$\log(L/L_{\mathrm{Edd}})$} & \colhead{SS73 $\tau_{\mathrm{0}}^{c}$} \\
\colhead{} & \colhead{deg} & \colhead{deg} & \colhead{} & \colhead{} & \colhead{\%} & \colhead{($\mathrm{erg\,s^{-1}}$)} & \colhead{($M_{\odot}$)} & \colhead{} & \colhead{days}
}
\startdata
267 & 212.80299 & 53.75199 & 0.588 & 19.6 & 19.4 & 44.41 & $7.42^{+0.17}_{-0.17}$ & -0.39 & 0.31  \\
300 & 214.92128 & 53.61379 & 0.646 & 19.5 & 18.2 & 44.87 & $7.6^{+0.17}_{-0.20}$ & -0.12 & 0.51 \\
399 & 212.63053 & 52.25938 & 0.608 & 20.1 & 23.6 & 44.22 & $7.91^{+0.16}_{-0.20}$ & -1.09 & 0.39 \\
551 & 212.94610 & 51.93883 & 0.681 & 21.5 & 10.3 & 44.33 & $6.95^{+0.19}_{-0.19}$ & -0.01 & 0.2\\
622 & 212.81328 & 51.86916 & 0.572 & 19.6 & 17.2 & 44.50 & $7.94^{+0.19}_{-0.16}$ & -0.83 & 0.5\\
634 & 212.89953 & 51.83459 & 0.651 & 20.8 & 13.2 & 44.06 & $7.56^{+0.26}_{-0.24}$ & -0.88 & 0.26\\
824 & 212.65879 & 52.00913 & 0.846 & 21.5 & 36.6 & 44.20 & $8.63^{+0.45}_{-0.45}$ & -1.82 & 0.67\\
840 & 214.18813 & 54.42799 & 0.244 & 18.6 & 50.0 & 43.49 & $7.93^{+0.21}_{-0.20}$ & -1.83 & 0.22\\
\enddata
\tablecomments{
\footnotesize
${}^a$ The fractional variability is the ratio of the root-mean-square (RMS) to average maximum likelihood flux calculated using the PrepSpec \citep{Shen2016a} software. The values reported here are computed from the existing 2014 - 2017 SDSS-RM PrepSpec lightcurves (http://star-www.st-and.ac.uk/$\sim$kdh1/pub/sdss/2017b/sdss.html), as reported in \citet{Shen2019a} for the first-year SDSS-RM data.
${}^b$ The black hole masses are drawn from \citet{Grier2017}.
${}^c$ The expected SS73 disk-size priors as computed from Equation~\ref{eq:tau0} (see Section~\ref{sec:t0}).
}
\end{deluxetable*}

\subsection{Cycle 25 HST UV Monitoring Campaign} \label{sec:cyc25}
The UV monitoring campaign was executed over two cycles of \hst\ observations (Cycle 25 \& 26)\footnote{The data can be obtained from the MAST archive at \dataset[10.17909/t9-2cc8-8s52]{https://doi.org/10.17909/t9-2cc8-8s52} and \dataset[10.17909/t9-bmkf-m360]{https://doi.org/10.17909/t9-bmkf-m360}.}. Five of the eight quasars (RM399, RM551, RM622, RM634, and RM824) were observed using \hst\ WFC3/UVIS F275W during 2018 March-May over 32 orbits, with an every-other-day cadence and a 63-day duration using four-point dither positions and 64-second subexposures. These targets were observed using the ``Drift-And-Shift," i.e., DASH observing design \citep{Momcheva2017}, dropping to gyro guiding after the first target (RM399) to avoid spending time on guide-star acquisition. Due to these targets' proximity, we fit all five quasars within a single visit. During the DASH sequence, our targets were observed in the following order: RM399, RM824, RM622, RM634, and RM551.
Out of the 32 visits, only visit 20 experienced a gyro failure. This visit was later compensated with an additional visit 33. Furthermore, target acquisition failed in visits three and four, which caused the entire DASH sequence to fail.

The typical drift under \hst\ gyro guiding is $\sim0.0015''/$ sec. However, the last targets (RM634 and RM551) in the DASH sequence often were dropped off the detector. The fourth (RM634) and fifth (RM551) targets were observed at $\sim$ 36 min and $\sim$45 min into each orbit. This indicates a drift $\gtrsim 0.007''/$sec for RM634 and $\gtrsim0.005''/$sec for RM551 that is significantly larger than the expected drift under star guiding. It is possible that our larger drift rate is caused by the larger slews between pointings while gyro guiding than the mosaic strategy of \citet{Momcheva2017}. We conclude that the drift due to dashing is at least 50\% of the time $\gtrsim 0.005''$/sec for our last two targets.

\subsection{Cycle 26 HST UV Monitoring Campaign} \label{sec:cyc26}
The UV monitoring campaign observed three other targets (RM267, RM300, and RM840) through \hst\ Cycle 26. These targets were observed during 2019 March-June with WFC3/UVIS F275W using non-DASHed observations over 40 orbits, with an every-other-day cadence and an 80-day duration with 52-second subexposures. All three targets fit in a single visit. However, the available roll angles affected the guide star availability, which resulted in RM840 being observed for only 33 visits. Removing RM840 from visits 33 to 40 increased the other two targets' available exposure time from 52 sec to 190 sec. \hst\ suffered from gyro failure in early 2019, and though it returned to science operations before our monitoring program began, this resulted in longer maneuvering time for target acquisition, which caused failures for two out of the three targets (RM300 and RM840) in visit 22 and failure for all three targets in visit 24.

\subsection{Ground Based Monitoring} \label{sec:ground}
The \hst\ UV monitoring program was accompanied by coordinated ground-based monitoring from the \textit{Liverpool Telescope} (\textit{LT}) and \textit{Las Cumbres Observatory Global Telescope Network} (\textit{LCOGT}). The LT observations were performed using the fully autonomous robotic systems with
the Spectral imager with a $10'\times10'$ field of view with a pixel scale $0\farcs152\, \rm{pixel}^{-1}$ (1$\times$1 binning) 
on the 2m telescope at the Haleakala site, and the Sinistro imager with a $26\arcmin \times 26\arcmin$ field of view and pixel scale of $0\farcs389$ (1$\times$1 binning) on the 1m telescopes at the McDonald site. The ground-based monitoring design is different between the two cycles of \hst\ monitoring. During Cycle~25, LT/IO:O (Infrared-Optical:\,Optical) observations provide $r$-band photometry, while LCOGT provides $r$-band and $z$-band photometry. During Cycle~26, we expanded the range of filters and used LT/IO:O to observe in $r$ and $z$-band while simultaneously observing in $g$ and $i$-band with LCOGT. Table~\ref{tab:table2} provides a short description of each telescope, duration, and number of contributed epochs for this study.

Our ground-based monitoring started before each \hst\ UV monitoring program and extended beyond the completion of UV monitoring observations. The extended duration allows the capturing of optical continuum variability, which typically has smaller amplitudes and longer timescales than UV variability. By extending the ground-based monitoring beyond the UV monitoring, we enable detection of longer lags, and high cadence allows detection of short lags. Our ground-based monitoring ideally has a daily cadence. However, the effective cadence due to weather loss was more sparsely sampled (with a mean of $\sim$1.5 days).

\startlongtable
\begin{deluxetable*}{lccccc|c}
\tablecaption{Summary of Observations \label{tab:table2}}
\tablehead{
\colhead{Observatory Name} & \colhead{Obs ID} & \colhead{Aperture} & \colhead{Observing Window} & \colhead{Filters} & \colhead{Epochs} & \colhead{Target RMID}\\}
\startdata
{\textit{HST UV  Monitoring}}&{}&{}&{}&{}&{}&{}\\
Hubble Cycle 25 & \hst\ 25 & DASH  & March - May (2018) & F275W & 32 & 399, 551, 622, 634, 824  \\
Hubble Cycle 26 & \hst\ 26 & non-DASH & March - June (2019) & F275W  & 40 & 267, 300, 840* \\
{\textit{Ground-based Optical Monitoring}}&{}&{}&{}&{}&{}&{}\\
Las Cumbres (McDonald)  & LCOGT  & 1.0 m  & Feb-May (2018) & $r$ & 54 & 551, 622, 824 \\
Las Cumbres (Haleakala) & LCOGT  & 2.0 m  & Feb-May (2018) & $z$ & 104 & 551, 622 \\
Liverpool Telescope  & LT  & 2.0 m & March-June (2018) & $r$ & 80 & 399, 551, 622, 634, 824 \\
Las Cumbres (McDonald)  & LCOGT  & 1.0 m & Jan-May (2019) & $g$ & 57-66 & 267, 300, 840 \\
Las Cumbres (McDonald)  & LCOGT  & 1.0 m & Jan-May (2019) & $i$ & 58-65 & 267, 300, 840 \\
Liverpool Telescope  & LT  & 2.0 m  & March-June (2019) & $r,z$ & 80 & 267, 300, 840 \\
\enddata
\tablecomments{
\footnotesize
*During Cycle~26, RM840 was observed for 33 orbits due to limited guide-star availability, see Section~\ref{sec:cyc26}.
}
\end{deluxetable*}

\section{Data Reduction} \label{sec:Reduction}

\subsection{\hst\ Cycle 25: DASH Observing Reductions}\label{sec:dash}
During Cycle 25, we adopted the ``Drift and Shift" (DASH) observing method, which reduces the overhead by using unguided, gyro-controlled exposures. This takes advantage of the available time in a single \hst\ visit by removing the requirement for a new guide-star acquisition between pointings. However, due to the lack of guidance sensor corrections, the telescope drift results in a smeared image by $0.''001$ - $0.''002$ per second \citep{Momcheva2017}. The DASH observing method has been successful in other IR wide-field studies such as COSMOS-DASH \citep{Mowla2019}. The WFC3/UVIS and WFC3/IR channels use the same pickoff mirror and Fine Guidance Sensor, and so we would expect both to experience the same telescope drift during gyro guiding, but DASH observing in the UV had not been directly tested until the current study.

Following the DASH observing mode during Cycle 25, we noticed that the smearing effect was far larger than expected in $\gtrsim90\%$ of visits. The automated reduction from the \texttt{astrodrizzle} pipeline \citep{Gonzaga2012} cannot identify the target from cosmic rays. Particularly, the target position shifts across the detector; this shift occasionally changed direction among the four subexposures. The smearing effect varies among the four subexposure dither positions and might extend across several pixels, with the fourth dither position generally being smeared the most. The shifted position of the target in each exposure usually caused it to be removed from the coadded images during cosmic-ray rejection in the standard reductions. We also tested other software for the automated reduction of the cosmic rays such as L.A. Cosmic \citep{vanDokkum2001}. However, we found these methods were only successful for our brightest target (RM622) but failed to identify the rest of the DASHed targets. To perform the photometric UV reductions, we first need to visually inspect to distinguish the target from the background cosmic rays and then perform the randomly smeared target's flux measurement.

\begin{figure}[t]
\centering
\includegraphics[width=60mm]{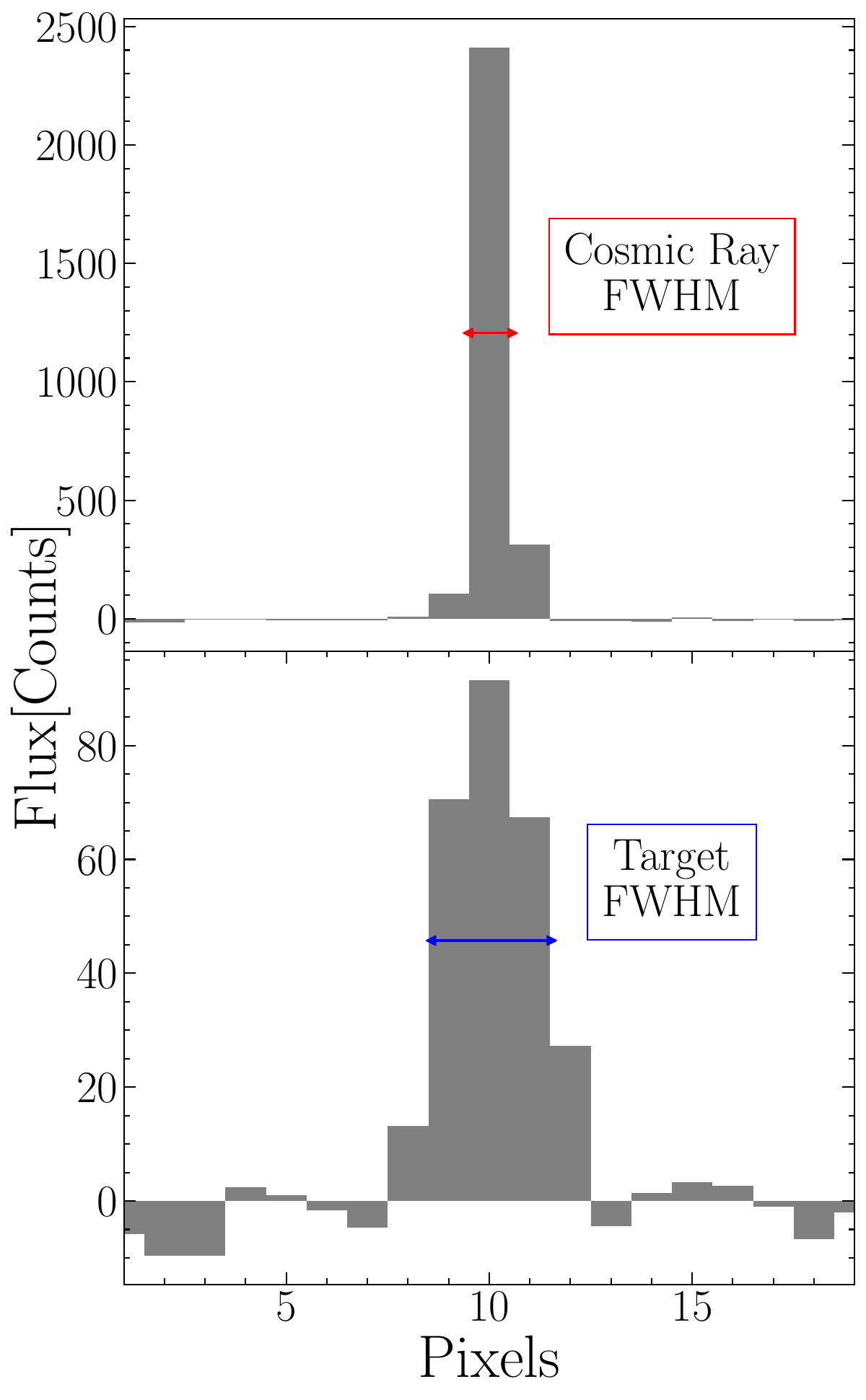}
 \caption{Comparison between the full width at half maximum (FWHM) for a cosmic ray and FWHM of the target: RM622. Cosmic rays typically appear with sharp edges on the image and thus have a narrower FWHM compared to a point-source quasar. We used this additional identification method during the visual inspection to distinguish our quasar targets from cosmic rays.}
\label{fig2}
\end{figure}

\begin{figure}[t]
\centering
\includegraphics[width=88mm]{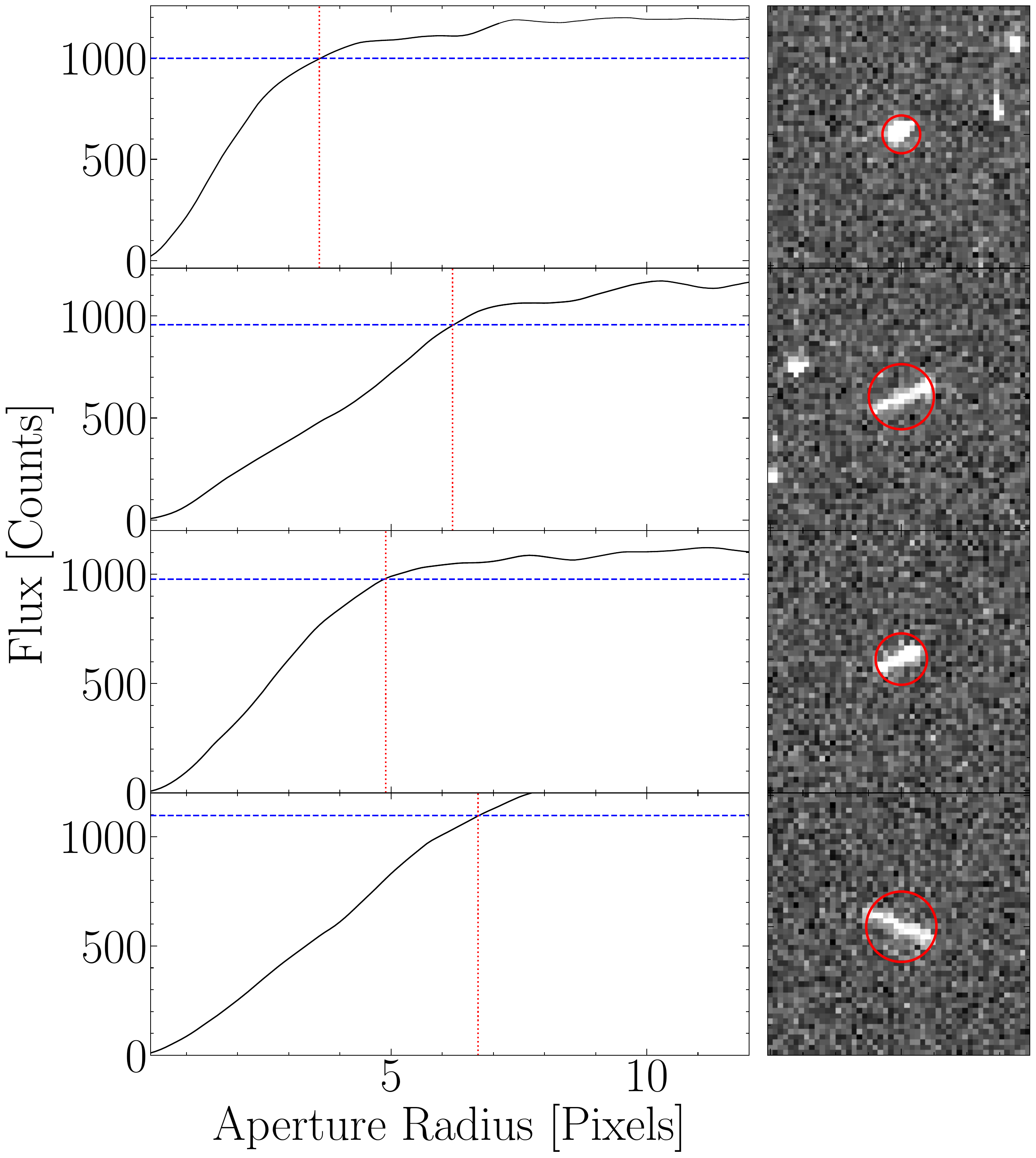}
 \caption{\textbf{Right:} An example of the four subexposure images for one of the quasars (RM622) observed with the DASH method. The target smearing varies in shape and direction, resulting in maximal smearing in the fourth image in this example. \textbf{Left:} Curves of growth for the flux as a function of aperture radius. We performed aperture photometry on a sequence of increasing circular radii ranging from 0 to 12 pixels. The chosen aperture size corresponds to 90\% of the target's flux saturation point (blue horizontal dashed line). The red vertical dotted line illustrates the final radius in pixels.}
\label{fig3}
\end{figure}

We used the calibrated, flat-fielded individual exposures (``FLT'' files) to locate our targets, visually identifying comparable objects appearing close to each other in successive subexposure images.
Our targets are less point-like and dimmer than the cosmic rays, resulting in a wider PSF with a lower maximum, as shown in Figure~\ref{fig2}. The average FWHM for RM622 is $\sim$2.5 pixels compared to the average cosmic-ray FWHM of 1.3 pixels.
This difference also translates to a visual distinction between cosmic rays and targets under extreme pixel distribution scales. Cosmic rays remain white and sharp, whereas our targets become gray and blurry, indicating a more gradual change in flux across the object and lower maximum flux. Additionally, with increased smearing, there is an increased contrast between the cosmic-ray and target PSF and appearance after adjusting the scale, making the most smeared objects the most conspicuously different among the comic rays.

The first target was always identified since it uses the standard star guiding. Among the DASH targets, we were only able to locate one of our targets (RM634) in five orbits, and therefore, we discontinued the analysis of this object. The remaining DASH quasars were identified in at least one subexposure image in 82\% of the visits.

We used the Astropy \texttt{photutils} \citep{Bradley2017} software package to perform aperture photometry. Identifying the optimum aperture for the flux extraction was complicated by the DASH observing method since the targets blurred into different shapes in each subexposure dither pointing. To account for this, we performed aperture photometry with circular apertures of increasing radii, $r_{\mathrm{aperture}}$. We adopted the circular aperture after comparing the signal-to-noise ratio from circular, rectangular, and elliptical apertures for different exposures (see the discussion below).
Testing radii on a range of $0< r_{\mathrm{aperture}} < 10$ pixels was sufficient for most targets but this was adjusted for more smeared targets to a range of $0 < r_{\mathrm{aperture}} < 20$ pixels. We estimated the local background within a circular annulus of $r_{\mathrm{inner}}$ equal to the maximum of the range for $r_{\mathrm{aperture}}$ and $r_{\mathrm{outer}} = r_{\mathrm{inner}}$ + 2 pixels. This results in an aperture mask for each subexposure dither pointing. We use the sigma clipped median estimator to obtain the local background. Using a median avoids outliers caused by the presence of high-flux cosmic rays in the annulus. The total background within each aperture is the local background times the circular aperture area.

We chose the aperture size to include 90\% of the object flux, illustrated in Figure \ref{fig3}. We performed this analysis on each subexposure dither pointing while visually inspecting each image.

We were able to obtain photometry from 70\% of the subexposure images. Photometry failed for targets that overlapped with cosmic rays and/or were too smeared or drifted off the detector (see Section~\ref{sec:cyc25}).

In some exposures the target and a cosmic ray overlapped, making the individual flux from each indistinguishable. This contamination is observed as a large, steep jump in the smoothly increasing target flux where the relatively high-flux, point-like cosmic ray is incorporated. Additionally, targets were sometimes excessively smeared and blended too much with the background (usually with $r_{\rm aperture}>15$ pixels). This level of smearing resulted in inaccurate and outlying low-flux measurements for a visit. For these targets, we tested rectangular and ellipse apertures. However, we found these exposures have a much smaller signal-to-noise ratio than the median target flux, and thus we rejected those subexposure dither pointings. Our method of examining the flux for a range of radii across the target allowed for clear identification and rejection of targets subject to both of these issues.

We compute UV flux uncertainties assuming a Poisson error distribution. We use the error array of the flat-fielded final pipeline outputs (FLTs) and compute the total flux uncertainty in each subexposure, $\sigma_{\rm{tot}}$, by adding the measurement uncertainties inside each aperture in quadrature, such that the $\sigma_{\rm{tot}}^2 = \sum_{\rm aperture} \sigma_{\rm error}^2$.

We use the reduced UV flux and flux-uncertainty measurements to compute the relative continuum UV lightcurve for four quasars (excluding RM634). We improved the final lightcurve quality by rejecting outlier flux measurements that were offset by more than three times the normalized median absolute deviation (NMAD e.g., \citealt{Maronna2006}). This excludes measurements affected by cosmic rays and/or large smearing.

\subsection{\hst\ Cycle26: UV Monitoring Reductions} \label{sec:cyc26_red}
For Cycle~26, we follow a similar reduction protocol as Cycle~25. Even though these observations are not performed using the DASH method, we adopt the Cycle~25 custom-reduction approach to remain consistent between our two sets of \hst\ observations. We use the flat-fielded subexposures at each dither positions and perform aperture photometry using the Astropy  \texttt{photutils} \citep{Bradley2017} software package. We test a sequence of 50 circular apertures in the range $1<r_{\rm aperture}<15$~pixels while estimating the local background from the sigma clipped median estimator (see Section~\ref{sec:dash}). We obtain the optimal aperture by computing the local maxima in the sum of flux over each $r_{\rm aperture}$. We compute the final target flux as $90\%$ of the maximum flux, computed from the sum of pixels in the optimal aperture from the background-subtracted, flat-fielded image. We estimate the flux uncertainties using the sum of error squares by placing the optimum aperture over the flat-fielded direct error outputs. We use these final flux and uncertainties to produce the relative photometric lightcurve for the three targets observed during Cycle 26.

After the custom reduction of the subexposure images was complete, we remove any bad measurements or outliers from the lightcurves. Some of the subexposures during Cycle~26 were affected by a persistent \hst\ gyro issue that caused the telescope to take much longer to acquire guide stars in between pointings. When this occurred, the telescope continued guide-star acquisition through a significant portion (up to 30 seconds) of the first exposure of the sequence.
This affected five out of 160 subexposures for RM267, 17 out of 160 subexposures for RM300, and 17 out of 132 subexposures for RM840. We removed these flagged subexposures from our final lightcurves. Similar to Cycle~25, we also excluded all subexposures that were offset by $>\,$3NMAD from the median lightcurve.

\subsection{Optical Monitoring Relative Photometry} \label{sec:optical}

To produce the relative photometric lightcurves for the ground-based observations, we select five standard stars for each telescope/field/pointing. We perform aperture photometry using the \texttt{photutils} \citep{Bradley2017} software package on the five standard stars of a magnitude similar to that of the quasars. Stars of similar brightness and color (compared to the target quasar) helps in identifying atmospheric effects distributed across the field of view for a uniformly-selected sample of reference stars and to avoid detector saturation (for bright references) and to avoid low signal-to-noise ratios (for faint references). Ideally, one is encouraged to utilize more references stars, however, here we chose five reference stars to remain consistent among all of our fields based on the availability of references. We extract the relative flux by calculating the ratio of the quasars' net integrated counts, $F_{\rm qso}$ to the sum of all the comparison stars, $F_{*}$:
\begin{equation}\label{eq1}
F_{\rm rel} = \frac{F_{\rm qso}}{\sum_{i}^{n} {F_{*}}_{i}},
\vspace{0.1cm}
\end{equation}
where the $i$ index indicates the ensemble's standard star. The aperture photometry is performed similarly to Section \ref{sec:dash} and \ref{sec:cyc26_red}; computing the rate of flux increase in the flat-fielded, sky background-subtracted image over 100 circular apertures in the range $1<r_{\rm aperture}<20$. We estimated the local sky background for each target from $r_{\rm{inner}} = r_{\rm{aperture}} + 3$~pixels to $r_{\rm{outer}} = r_{\rm{aperture}} + 6$~pixels. We find the optimal aperture for each quasar per observation using the local maxima of the flux increase over the aperture sequence. We extract the relative star lightcurves and, after visual inspection, substitute any variable star lightcurve with non-variable replacements.

\begin{figure}[t]
\centering
\includegraphics[width=88mm]{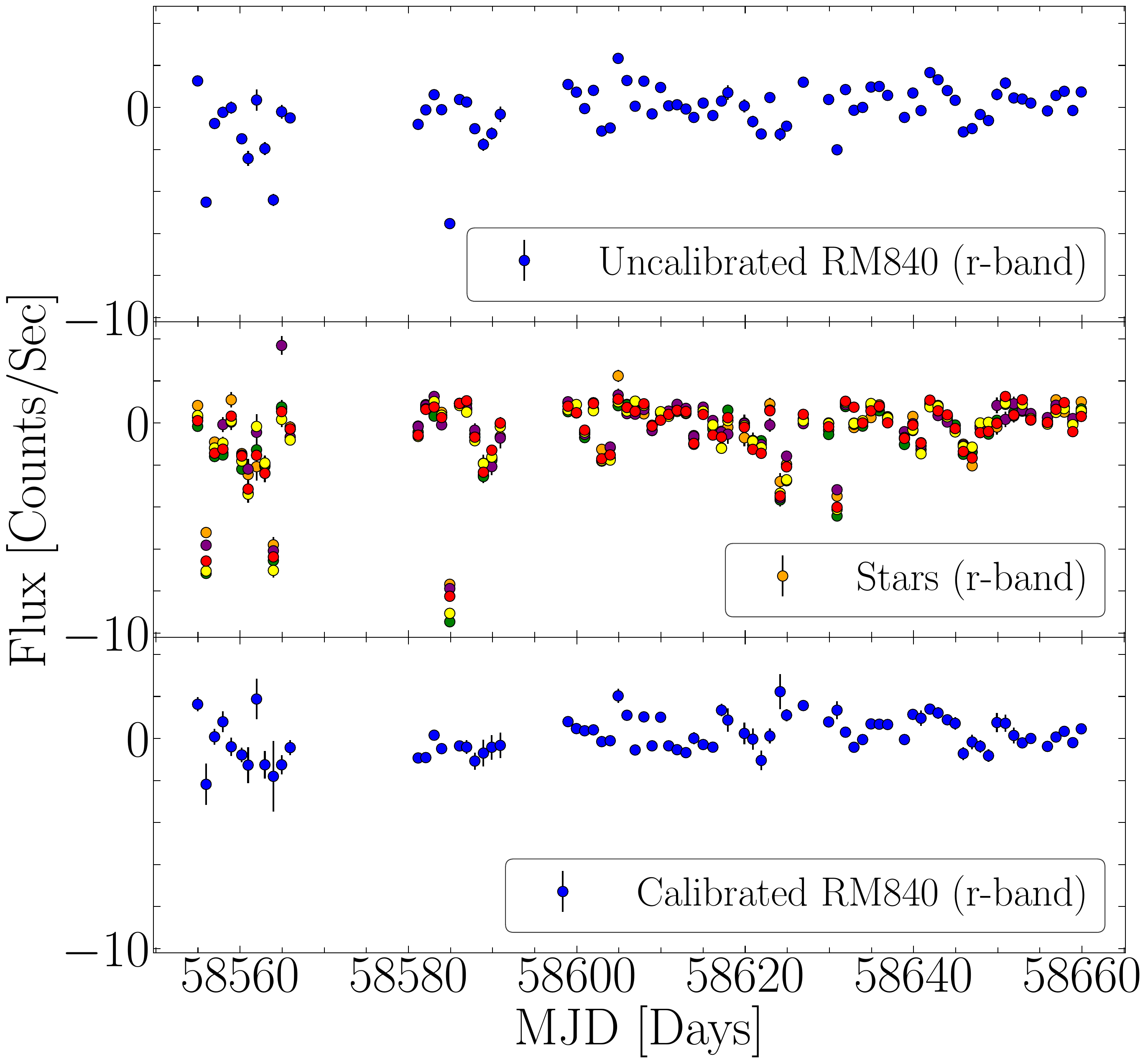}
 \caption{Uncalibrated r-band lightcurves for the quasar RM840 (top) and a standard star in the same images (middle), and the final r-band relative flux lightcurve for the quasar (bottom). For each quasar we select five standard stars to control for weather and instrumental effects on the raw, uncalibrated lightcurves and produce calibrated relative photometry of quasar variability (see Section~\ref{sec:optical}). After we produce the relative flux lightcurve, any outlier epochs that are offset by $>3\times$NMAD are rejected from the final lightcurve. For clarity, each of the reference stars and quasar lightcurves are displayed with mean of zero and errorbars are normalized to NMAD of 1.}
\label{fig4}
\end{figure}

We compute the flux uncertainties assuming Poissonian error for each aperture. We propagate the uncertainties from all apertures to derive the error in relative flux measurements. First, the uncertainty from each aperture photometry measurement of each standard star is combined in quadrature to give the total star ensemble uncertainty:
\begin{equation}\label{eq2}
    \sigma_{*\rm \, ensemble} = \sqrt{\sum_{i}^{n}{\sigma^2_{*}}_{i}},
\end{equation}
where $\sigma_{*}$ is the uncertainty of each star in the ensemble, and index $i$ is the number of  standard stars. The propagated relative flux uncertainty is then give by:
\begin{equation}\label{eq3}
    \sigma_{\rm rel} = \frac{F_{\rm qso}}{F_{*\rm \, ensemble}} \sqrt{\frac{{\sigma}_{\rm qso}^2}{F_{\rm qso}^2} + \frac{\sigma_{*\rm \, ensemble}^2}{F_{*\rm \, ensemble}^2}},
\end{equation}
where $F_{\rm qso}$ is the net integrated counts per second in the quasar aperture, $F_*$ is the sum of the net integrated counts per second in the ensemble of standard stars, $\sigma_{\rm qso}$ is the uncertainty in the quasar aperture, and $\sigma_{\rm * ensemble}$ is the uncertainty of the standard-star ensemble from Equation~\ref{eq2}. We compute the relative flux and flux uncertainties from Equation~\ref{eq1} and Equation~\ref{eq3} respectively, using individual apertures for each standard star, quasar, filter, and field to produce all ground-based lightcurves. See Figure~\ref{fig4} for an example of this comparison between the raw and relative lightcurve. We also experimented with other photometric extraction technique, including difference imaging as implemented by Danida \citep{Bramich2008} but this led to similar SNR. We test the impact of the optical lightcurve SNR on the measured lags in the appendix.

We used the weighted average between repeated exposures within a night and computed the final lightcurves. We additionally removed any measurements that were $\times$3NMAD offset from the median of the entire lightcurve.

An examination of the lightcurve variability between epochs reveals that custom relative photometry reduction may introduce over-estimated errors that mask the underlying flux variability.
We follow the procedure outlined in \citet{Grier2017,Grier2019, Homayouni2020}, and apply error rescaling by using the lightcurve intercalibration step described below.

\section{Time Series Analysis}\label{sec:timeseries}

\subsection{Lightcurve Intercalibration} \label{sec:cream_intercalibration}
Supplementary ground-based optical monitoring with LCOGT and LT observations provide sufficient cadence to generate well-sampled lightcurves over multiple bands ($g,r,i,z$). To perform the lag analysis, we must merge observations from different telescope sites with different seeing, filter throughputs, and local sky backgrounds on the same flux scale by intercalibrating each target's lightcurve.

We perform this intercalibration by using the \cream\ \citep[Continuum REprocessing AGN Markov chain Monte Carlo;][]{Starkey2016, Starkey2017} merging feature to combine lightcurves at the same wavelength but taken from different telescopes. \cream\ uses a Markov chain Monte Carlo (MCMC) approach to fit a normalized driving lamp-post model to continuum lightcurve and the accretion disk response function, and infer the posterior probability distribution for the disk temperature $T_1$, temperature slope $\alpha$, and disk inclination $i$ \citep{Starkey2016}. During this process, \cream\ performs the merging by transforming the old lightcurve $f_{\mathrm{j,old}}(\lambda,t)$ to the new lightcurve using $f_{\mathrm{j,new}}(\lambda,t)$ following Equation~3 in \citet{Grier2017},
\begin{equation}\label{Eq:CREAM}
    f_{j,new}(\lambda,t) = (f_{j,old}(\lambda,t) - \bar{F_j}) \frac{\Delta F_{\rm REF}}{\Delta F_j} + \bar{F}_{\rm REF}
\end{equation}
Where $\bar{F_j}$ and $\Delta F_j$ are the mean and standard deviation of the $j$ lightcurve respectively, which will be mapped onto the reference lightcurve with mean and standard deviation $\bar{F}_{\rm REF}$ and $\Delta F_{\rm REF}$ using Equation \ref{Eq:CREAM}; also, \cream\ can adjust the underestimated (or overestimated) error bars by adding two parameters to model inaccurate error bars. For each telescope/filter lightcurve, the rescaled lightcurve is computed using
\begin{equation}\label{Eq:Cream_error}
    \sigma_{ij}^2 = (S_j\,\sigma_{old,\,ij})^2 + V_j
\end{equation}
where $i$ indicates the number of data points for each telescope/reference lightcurve and $V_j$ is the extra variance, and $S_j$ is the scale-factor parameter. The likelihood function for each telescope/filter lightcurve penalizes large values of $V_j$ and $S_j$.

\cream\ simultaneously fits the offset and rescaling parameters we use to inter-calibrate observations from different sites, and rescales the overestimated lightcurve uncertainties while also inferring the lamp-post lightcurve that drives the continuum variability. This paper's entire time series analysis is performed using the rescaled and intercalibrated lightcurves generated from \cream.

\subsection{Lag Identification}\label{sec:method}
We adopt two time-series analysis methods for measuring reverberation lags: \jav\ \citep{Zu2011} and \cream\ \citep[also see Section~\ref{sec:cream_intercalibration}]{Starkey2016}. Similar to \cream, \jav\ \citep{Zu2011} uses a damped random walk (DRW) model to describe the stochastic variability of the quasar lightcurves. Even though the DRW model may be an incomplete description for quasars on short timescales \citep{Mushotzky2011, Kozlowski2016}, studies have shown that DRW model still provides a flexible approach to accurately measuring lags \citep{Li2019, Read2020} and a reasonable fit to observations of quasar variability on the timescales of our monitoring program (days to weeks) \citep{Kelly2009, MacLeod2010, MacLeod2012, Kozlowski2016}.

\jav\ uses a Markov chain Monte Carlo approach using a maximum likelihood method to fit a DRW model to the UV and optical continuum lightcurves, assuming that the local accretion-disk response is a top-hat function and the reverberating lightcurve model is the smoothed, scaled, and shifted version of the UV continuum lightcurve.

We allow the DRW amplitude to be a free parameter but fix the DRW damping timescale to 100 days. Our campaign duration ($\sim$80 days) is much smaller than the typical damping time scale of a quasar ($\sim$1500 days in observed-frame, see \citealp{Kelly2009, MacLeod2012}). Thus, the damping timescale's exact value does not matter, so long as it is longer than the campaign's duration (the lightcurves are effectively modeled as a red-noise random walk with minimal damping). We also tested damping timescales of 200 and 300 days and found no significant differences in the measured lags, as also investigated by \citet{Yu2020}.

The optical lightcurve response is parameterized as a top-hat transfer function, assuming a lag and scale factor with a free parameter. The top-hat transfer function in \jav\ is a simplification of the actual transfer function from the accretion disk, which may be extended with a long tail at large lags and affect the \jav\ measurements \citep{Starkey2016}. This means that the \jav\ measurements may be underestimates of the actual mean disk lags. However, the top-hat transfer function is commonly adopted in other works \citep{JiangGreen2017, Yu2020, Homayouni2019} and so we adopt the simple top-hat transfer function here to provide consistency for comparison of our lag measurements with previous work. We fix the transfer-function width to be 0.5~days, which is sufficiently short compared to the expected lag ($1<\tau<14$~days). We tested a wide range of transfer function-widths 0.1-10 days, which affected the convergence of the MCMC chain in \jav\ but did not significantly affect the best-fit lag (so long as the \jav\ chain still converged). We adopt a lag search range of $\pm$45 days (Cycle~25) and $\pm$60 days (Cycle~26), chosen to be $\sim \times 2/3$ of the $\sim$60 and $\sim$80~day monitoring duration. All of our final measured lags (see Table~\ref{tab:table4}) are significantly shorter than these search ranges. \jav\ returns a lag-posterior distribution from 62500 MCMC simulations, which are used to compute \jav\ lag, $\tau_{jav}$, and its uncertainty. Among the targets in our sample, \jav\ was unable to obtain a continuum model for RM551 using the final \cream-rescaled error bars and successfully produced the DRW lightcurves only after we further rescaled the error bars by $\sim$ 80\% (see Figure set for RM551).

We also use the \cream\ \texttt{Python} wrapper, \pycecream\footnote{https://github.com/dstarkey23/pycecream} to infer accretion-disk lags in addition to the intercalibrating lightcurves (see Section \ref{sec:cream_intercalibration}). We probe lags of $\pm$ 50 days to obtain \cream\ lag posterior distributions.

The \jav/\cream\ MCMC posterior-lag distributions may have a few ancillary peaks that accompany a primary peak. To identify the reverberating lag from lag posterior distributions, we smooth each posterior by a Gaussian filter with a 3-day $\sigma$ (the width of the smoothing was determined by visual inspection). We then identify the primary peak of the posterior distribution from the peak with the largest area and treat the smaller-area peaks as insignificant lags. The final lag, $\tau$, and the lag 1$\sigma$ uncertainty is computed from the median, the 16th and the 84th percentiles of the posteriors in the primary peak.

\begin{figure*}
\centering
\includegraphics[width=170mm]{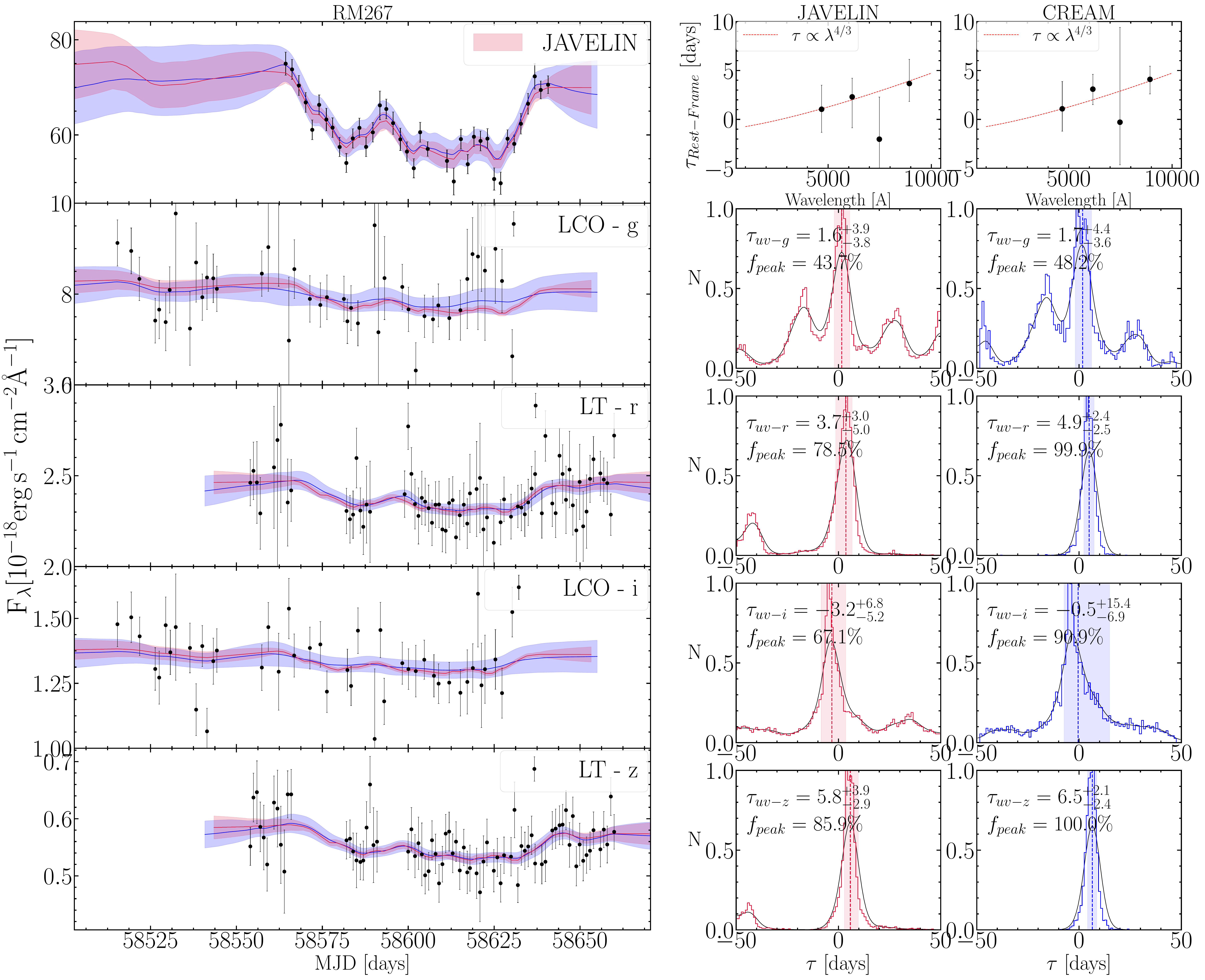}
 \caption{\textbf{Left:} Continuum F275W lightcurve (top) and the optical $griz$ lightcurves are shown respectively from top to bottom for RM267. For each lightcurve, the best-fit DRW model and uncertainty are displayed in the shaded region for \jav\ (red) and \cream\ (blue). The F275W lightcurves displayed here show the weighted-average observations for the four-point dither positions. \textbf{Top Right:} The rest-frame lags (with respect to the UV) versus optical wavelength, with a $\tau\propto\lambda^{4/3}$ relation shown as red line. The bottom panels show the UV-optical lag posterior distribution for \jav\ (left/red) and \cream\ (right/blue) and the final observed-frame lag is displayed for each method. For each lag posterior distribution, the final observed-frame lag and the 16th/84th percentile error bar are illustrated by the vertical dashed line and the shaded region, respectively. The complete figure set (7 images) is available for all targets.}
\label{fig5a}
\end{figure*}

For each target, we measure the inter-band lags between the F275 W filter, $\lambda_{\mathrm{pivot, uv}}$~=~2704 \AA, and optical $g,r,i,z$ bands at $\lambda_{\mathrm{cent}}$~=~4686, 6166, 7480, and 8932 \AA\ respectively. Figure~\ref{fig5a} (see also the complete figure set similar to Figure~\ref{fig5a}) illustrates each target's UV and optical lightcurve, \jav\, and \cream\ lag posterior distributions, and the rest-frame lag compared to SS73 wavelength scaling. In this work, we use both methods to perform the accretion-disk analysis. This enables comparison of both methods' lag results considering our medium-quality lightcurves following recent comparisons of lag methodologies for survey-quality RM observations \citep{Li2019} and continuum RM accretion disk lag methods \citep{Chan2020} and their implications for a statistical approach to modeling the disk structure.

Using either method, we find that \jav\ and \cream\ lags generally produce consistent lag posteriors. There are three lag posteriors where the final lags are inconsistent; UV-$z$ in RM551 and UV-$r$ and UV-$z$ in RM622 (see the Figure set). In all these cases, \jav\ detects a larger negative lag compared to \cream. This may be due to larger lightcurve uncertainties where \jav\ is originally unable to fit a DRW without any custom error bar rescaling (see the discussion earlier in this section). It also might indicate that a top-hat is an over-simplified assumption for the disk-response function in this quasar. \cream\ uses a disk-response function that rises rapidly to a peak and has a long tail toward large lags and is likely a better description of the disk response.

We find that in most cases the longer wavelength continuum variation lags behind those at shorter wavelengths, as expected for reverberation in a lamp-post model. However, the increasing lag with wavelength has exceptions in the $i$-band filter. For targets where we have multi-band observations, we see that $i$-band observed-frame lag is occasionally much shorter, $\tau = -3.2_{-5.2}^{+6.8}$~days (RM267; see Figure~\ref{fig5a}), or much longer, $\tau = 23.4_{-7.8}^{+6.6}$~days  and $\tau = 35.6_{-5.1}^{+5.9}$~days (RM300 and RM840, respectively) than lags in other filters. In addition, the $g$-band lag for RM840 is much larger than expected, $\tau = 32.1_{-4.9}^{+5.7}$~days (see figure set). These larger lags could be due to effects from the emission lines in the  BLR, contributions from the iron pseudo-continuum or the diffuse Balmer continuum \citep{Korista2001, Korista2019, Lawther2018}. We will discuss these contributing factors and other lag-measurement reliability components in Section \ref{sec:blr}. It is more difficult to assess the trend of larger lags at longer wavelengths for those targets that were observed as part of Cycle~25 due to lag aliasing issues and larger uncertainties. That said, we find that the \cream\ lags in RM551 are in agreement with a larger lag at longer wavelengths.

\subsection{BLR Contamination} \label{sec:blr}

One plausible scenario for longer continuum lags is that these lags may be originating in the BLR, where the BLR or diffuse emission significantly contributes to the variability measured in the photometric filter, and the typically longer BLR lags make the measured photometric lag larger than expected solely from continuum lags.

Some investigation of these larger lags reports that the BLR emission is responsible. \citet{Fausnaugh2016, Cackett2018} found evidence for nebular BLR contamination in specific filters, and \citet{CheloucheZucker2013, Chelouche2013} found this to be a widespread phenomenon. Considering the wavelength range of our observations and our target redshifts, we investigated the effect of BLR contamination from prominent BLR emission lines: \Lya\,$\lambda$1215,  \CIV\,$\lambda$1549, \CIII\,$\lambda$1909, \MgII\,$\lambda$2800, \HeII\,$\lambda$4687, \Hb\,$\lambda$4861, and \Ha\,$\lambda$6563.
To compute the BLR contribution, we examine whether an emission line falls in the filter bounds in the quasar's observed-frame. If so, we then use the ratio of emission-line equivalent width, EW$_{\rm{line, rms}}$ from Prepspec outputs to the overlapping filter width. \citet{Shen2019a} provides a full description of PrepSpec fits applied to first-year SDSS-RM observations\footnote{The PrepSpec outputs from 2014-2017 SDSS-RM observations are available at http://star-www.st-and.ac.uk/$\sim$kdh1/pub/sdss/2017b/sdss.html} (Horne et al. in prep.). We obtain the fractional BLR contamination by multiplying this ratio by the root-mean-square (RMS) variability of the emission line and nearby continuum ($\lambda1700$, $ \lambda3000$, $\lambda5100$). Table~\ref{tab:table3} summarizes the contribution from the BLR emission line contribution for all objects in our sample. Only the $uv-i$ lag in RM840 exhibits a maximum 13\% contribution from the \Ha\ emission line, which we reject by choosing a BLR contamination rejection threshold of 10\%. We note that the \Ha\ lag reported for this object \citep{Grier2017} is only $13.2_{-3}^{+2.9}$ days, which contradicts a simple BLR contamination by \Ha. The BLR contamination in the rest of our targets falls well below the 10\% contamination limit.

\startlongtable
\begin{deluxetable}{ccccc}
\tablecaption{BLR Contamination\label{tab:table3}}
\tablehead{
\colhead{RMID} & \multicolumn{4}{c}{Emission Line Contamination (\%)}\\
\colhead{} & \colhead{$g$-band} & \colhead{$r$-band} & \colhead{$i$-band} & \colhead{$z$-band}
}
\startdata
RM267 & 0.7 & - & 4.3& -\\
RM300 & 0.1 & - & 0.5& -\\
RM399 & 0.8 & - & 2.4& -\\
RM551 & 0.2 & - & 7.4& -\\
RM622 & 0.7 & - & 3.6& -\\
RM634 & 0.2 & - & 0.5& -\\
RM824 & 1.5 & - & - & 6.8\\
RM840 & -  & 1.8 &13.0 & -\\
\enddata
\end{deluxetable}

In addition to the emission-line BLR contamination, some quasars may have significant contributions from diffuse continuum emission from the BLR clouds \citep{Korista2001}. The contribution from this variable diffuse emission originates in the BLR, at larger radii than the continuum variability of the accretion-disk. \citet{Korista2001, Korista2019} have claimed that the diffuse Balmer continuum significantly affects the interband continuum lags observed in NGC7469 \citep{Wanders1997, Collier1998, Kriss2000, Pahari2020}. This effect was particularly apparent near the Balmer jump 3646\,\AA\ in the lag spectrum of NGC4593 \citep{Cackett2018} and also in other studies of local AGNs \citep{Edelson2015, Fausnaugh2016, Edelson2017, Edelson2019, Cackett2020}. There are two main contributors to the diffuse Balmer continuum. The first source is emission from free-bound transitions (recombination continuum), which affects wavelengths bluer than Balmer edge. The second contributing factor is blended high-order bound-bound transitions, which results in a diffuse Balmer forest red-ward of the Balmer edge. This effect could explain the large UV-$g$ and UV-$i$ lags that are $\gtrsim$~10 days (observed-frame) and overlap with 3646~\,\AA. We thus exclude any observed-frame lags $>10$ days in filters that overlap with rest-frame $\lambda$\,3500~-~3900\,\AA.

Furthermore, a plethora of weak emission lines from many thousands \FeII\ transitions in the BLR form a pseudo-continuum that spans UV to near-infrared wavelengths \citep{Vestergaard2001, Bruhweiler2008}. This slowly-varying \FeII\ pseudo-continuum introduces uncertainty in the true continuum variability \citep{Kuehn2008}.
We thus exclude any outlier lags that overlap the \FeII\ complex at $\lambda$\,4434~-~4684\,\AA\ \citep{Boroson1992} or $\lambda$\,5100~-~5477\,\AA\ \citep{VandenBerk2001}. The UV \FeII\ pseudo-continuum at $\lambda$\,1250~-~3090 \AA\ (rest-frame) \citep{Vestergaard2001} generally has little effect on the continuum fluxes in our observed-frame filters. Typical \FeII\ equivalent widths are small ($<$ 50 \AA), and so we anticipate minimal contribution from iron emission. We reject outlier measurements that fall within these windows and have rest-frame lags that are too large ($>$10~days or $<$-10~days); the rejected outlier lags include four measurements: $\tau_{uv-i\rm{\,(RM300)}} = 14.2_{-4.7}^{+3.9}$,  $\tau_{uv-r\rm{\,(RM399)}} = 15.9_{-4.4}^{+4.7}$,  $\tau_{uv-z\rm{\,(RM551)}} = -16.3_{-4.9}^{+4.4}$, and  $\tau_{uv-g\rm{\,(RM840)}} = 25.7_{-3.9}^{+4.6}$ using \jav\ and $\tau_{uv-i\rm{\,(RM300)}} = 17.1_{-5.1}^{+4.9}$,  $\tau_{uv-r\rm{\,(RM399)}} = 14.4_{-3.6}^{+4.1}$,  $\tau_{uv-z\rm{\,(RM551)}} = 14.6_{-4.6}^{+4.5}$, and  $\tau_{uv-g\rm{\,(RM840)}} = 25.9_{-2.7}^{+3.1}$ using \cream. These rejected lags along with other insignificant lags (see Section~\ref{sec:reliability} for individual target discussion) are shown with open symbols in Figure~\ref{fig6}. The diffuse Balmer and \FeII\ pseudo-continuum windows are also shown as gray-shaded regions in Figure~\ref{fig7}.
\subsection{Lag Reliability} \label{sec:reliability}
The lag posterior distribution from \jav\ or \cream\ has a primary peak, which corresponds to a flux-weighted mean radius for emission in the bandpass. This primary lag is identified as the smoothed lag posterior region between local minima with the largest area. This primary peak is often accompanied by less-significant peaks, which may be interpreted as alias lag solutions. To ensure that the final reported lags are statistically meaningful, we require ``reliable" lags as those containing a minimum of 50\% of the lag posteriors samples within the primary peak, i.e., $f_{\rm peak} > 50\%$, following a similar approach to \citet{Grier2017, Homayouni2019}. The $f_{\rm peak}$ requirement ensures a reliable lag solution and removes cases with many alias lags in the posterior.

\begin{figure*}[tt]
\centering
\includegraphics[width=180mm]{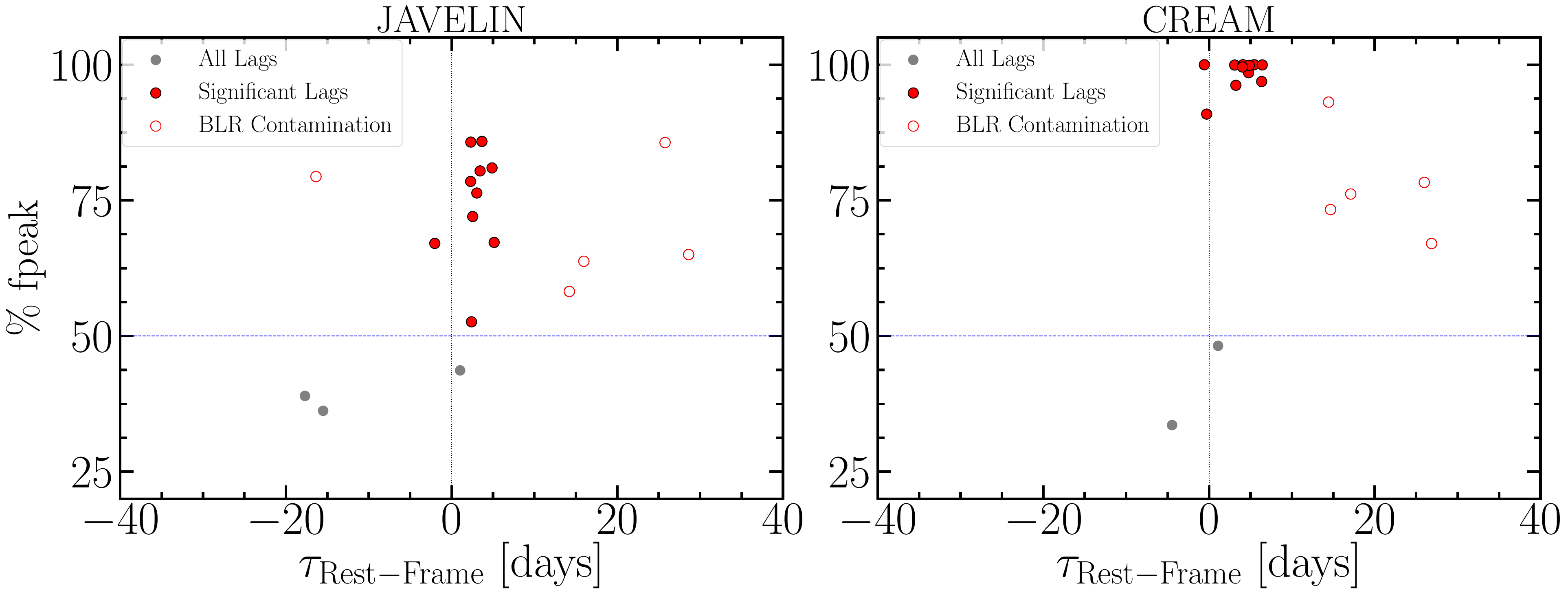}
 \caption{Lag significance criteria for \jav\ (left) and \cream\ (right) methods. A lag is significant if its peak (between local minima) includes at least 50\% of the lag posterior samples and has $<$10\% contribution from the prominent emission lines. If a prominent broad emission line falls in any of the filter ranges and adds significantly to the continuum variability, the lag is considered contaminated and removed from the significant lag measurements. In addition, the diffuse Balmer continuum at 3646\,\AA, and the \FeII\ pseudo-continuum at $\sim \lambda\, 4434-4684$\,\AA\ and $\sim\lambda\,5100-5477$\,\AA\ may contribute to the excess of larger lags in these regions (see Section~\ref{sec:blr}). Red open symbols show outlier lags ($>$ 10~days  or $<$-10~days in rest-frame) that may be affected by diffuse BLR contamination despite having $f_{\rm peak}>$50\%.}
\label{fig6}
\end{figure*}

Figure \ref{fig6} shows the lag-measurement results for all of the inter-band lags for our targets. Considering the different observation design and optical filter coverage during Cycle 25 and Cycle 26, we cover 18 inter-band lag measurements. The lag-significance criteria are shown in each panel. Out of the 18 inter-band lags distributed among 7 targets, \jav\ finds 10 significant lags and \cream\ finds 11 significant lags. Table~\ref{tab:table4} reports our final significant lag measurements.

We review each target's lag measurement (for either \jav\ / \cream\ method). We discuss the lags measured or rejected for each quasar in detail below.
\begin{itemize}
    \item[] \textbf{RM267}: For this target, the reliable lag measurements are limited to UV-$r$, UV-$i$, and UV-$z$ lag measurements.

    The UV-$g$ lag has $f_{\rm peak}<$ 50\% and is considered insignificant.

    The UV-$r$ lag overlaps with the diffuse Balmer continuum (at $\sim$ 3882\,\AA). However, the rest-frame lag is relatively short, $\tau_{\rm{rest-frame (jav)}} = 2.3_{-3.2}^{+1.9}$ days, and therefore it is unlikely to be significantly contaminated by diffuse Balmer emission from the BLR.

    \item[] \textbf{RM300}: For this target, the only reliable lags are UV-$g$, UV-$r$, and UV-$z$ lag measurements.

    The UV-$r$ lag falls in the diffuse Balmer continuum  window at $\sim$3746\,\AA. However, the lag is short and so is unlikely to be significantly affected by the diffuse Balmer continuum.

    The UV-$i$ lag, on the other hand, overlaps with the \FeII\ pseudo-continuum at 4544 \,\rm\AA\ with rest-frame lag $\tau_{uv-i} = 14.2_{-4.7}^{+3.9}$ days and is therefore rejected from our final reliable lag sample.

    The UV-$z$ lag also overlaps with the \FeII\ pseudo-continuum ($\sim$5426\,\AA) but is short ($2.4_{-2.4}^{+3.0}$) and so is consistent with continuum variability dominating the lag rather than diffuse BLR contamination.

    \item[] \textbf{RM399}: The UV-$r$ lag is the only significant lag measurement for this target, and it overlaps with the diffuse Balmer continuum at $\sim$ 3834\AA. The size of this lag $\tau_{uv-r} = 15.9_{-4.4}^{+4.7}$~days is likely affected by the diffuse Balmer emission. We reject this lag measurement from our final lag sample.

    \item[] \textbf{RM551}: The only reliable lag for this target is the UV-$r$ lag from \jav.
    The UV-$r$ band at 3668\AA\ falls in the the diffuse Balmer window; however, the rest-frame lag is too short, $\tau_{uv-r} = 3.5_{-3.6}^{+3.7}$ using \jav\ and $\tau_{uv-r} = 6.4_{-4.4}^{+4.5}$ using \cream, to be significantly affected by the diffuse Balmer emission.

    The UV-$z$ lag overlaps with the diffuse \FeII\ pseudo-continuum at 5313 \AA. The reported lag is an outlier from both lag methods (a negative lag using \jav\ and a large positive lag using \cream).

    \item[] \textbf{RM622}: The only significant lag for this target is the \cream\ UV-$r$ lag.
    For this target, \jav\ reports a UV-$r$ lag with $f_{\rm peak}<$ 50\%. But the UV-$r$ lag using \cream\ is significant and does not overlap with any diffuse BLR emission windows.

    The UV-$z$ lag has $f_{\rm peak}<$ 50\% using either method and is therefore insignificant.

    \item[] \textbf{RM824}: The UV-$r$ is the only lag measurement for this target and is considered a significant lag with no overlap in the contributing diffuse BLR emission windows.

    \item[] \textbf{RM840}:  The UV-$r$ and UV-$z$ lags for this target are considered significant with no contribution from the diffuse BLR windows.
    The UV-$g$ at rest-frame $\lambda$~3766\AA\ lag falls in the diffuse Balmer continuum window. Considering the reported rest-frame lag of $\tau_{uv-g} = 25.7_{-3.9}^{+4.6}$, it is likely that this lag is significantly affected by the diffuse Blamer emission.

    The UV-$i$ lag in this target has a significant contribution from the BLR emission line (\Ha) and is therefore rejected from our final lag sample.
\end{itemize}

\begin{figure*}[t]
\centering
\includegraphics[width=180mm]{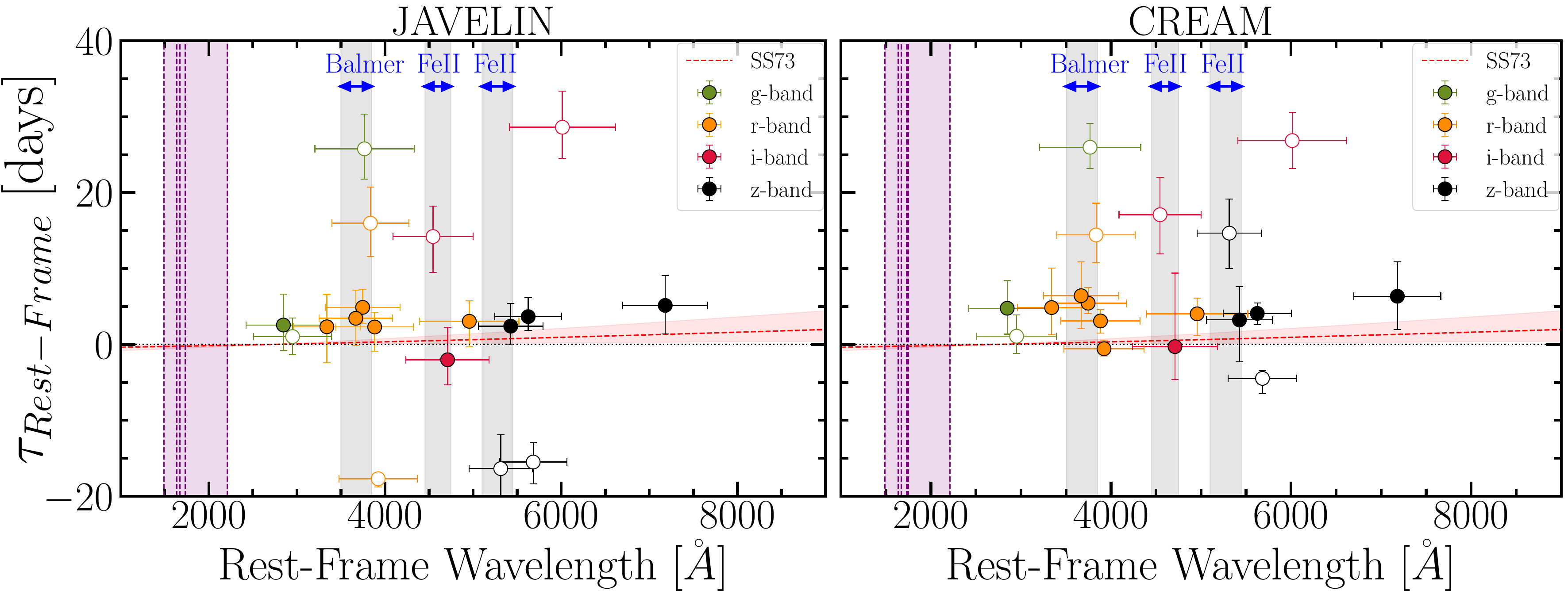}
 \caption{Rest-frame lag as a function of wavelength using \jav\ (left) and \cream\ (right). The colored symbols show the UV-$g$ (green), UV-$r$ (orange), UV-$i$ (red), and UV-$z$ (black). All the lags are computed with respect to the \textit{HST} UVIS F275W with a normalization wavelength of 2700\AA. The purple shaded region indicates the range of rest-frame F275W probed by 2750 \AA\ observed-frame, where the purple dashed line correspond to each quasar in our sample. The red dashed line displays the SS73 model for the mean \Mbh\ and mean \Mdot\ of the sample of significant lags, and the shaded region around the red dashed line illustrates the minimum and maximum SS73 disk size, computed from our sample's minimum and maximum in \Mbh\ and \Mdot. The shaded gray regions illustrate the wavelength regions potentially affected by the Balmer diffuse continuum and the \FeII\ diffuse continuum. We reject outlier ($|\tau|>10$~days) lags that overlap with these windows, which results in the rejection of $\tau_{uv-i}$ in RM300,  $\tau_{uv-r}$ in RM399, $\tau_{uv-z}$ in RM551, and  $\tau_{uv-g}$ in RM840. We have also rejected \textbf{$\tau_{uv-i}$} in RM840 due to high (13\%) BLR emission-line contamination from \Ha. All the rejected lags (see Sections \ref{sec:blr} and \ref{sec:reliability}) are illustrated with open symbols.}
\label{fig7}
\end{figure*}

Figure~\ref{fig7} illustrates all of our lag measurements, and Table~\ref{tab:table4} presents a summary of our final significant UV-optical continuum lags for our targets. For the remainder of this analysis, we remove the insignificant lags from our analysis and only use our reliable measurements. We perform accretion-disk structure analysis in Section \ref{sec:discussion}, interpreting the observations in comparison to the SS73 disk expectation.

\section{Discussion}\label{sec:discussion}
One of our UV-monitoring campaign's main goals is to use the UV-optical time delays to study inner-disk structure as a function of \Mbh\ and accretion rate. Our sample's redshift range $0.24\,<\,z\,<\,0.85$ translates to wavelength-dependent continuum lags that probe $\lambda\,$2847~-~7180\AA\ in the quasar rest-frame. We note that our significant lags per target include three inter-band lag measurements at best (see Table~4) as described in detail later in this section, which is not sufficient to constrain accretion disk parameters for each target individually. We combine the significant lag measurements for our targets (see Table~4) and use a Bayesian approach to fit an accretion-disk model parameterized as:
\begin{equation}\label{eq:tau_ss73}
    \tau_{\mathrm{opt}} - \tau_{\mathrm{uv}} = \tau_0\Bigg[\big(\frac{\lambda_{\mathrm{opt}}}{2700\mathrm{\AA}}\big)^{\beta} - \big(\frac{\mathrm{\lambda_{uv}}}{2700\mathrm{\AA}}\big)^{\beta}\Bigg]
\end{equation}
Here $\tau$ is the rest-frame lag, $\lambda_{\mathrm{uv}}$ is the rest-frame UV reference wavelength, and $\lambda_{\mathrm{opt}}$ corresponds to rest-frame optical wavelengths. In the ``standard'' optically thick, geometrically thin disk model \citep{SS1973}, $\beta=4/3$ and the disk normalization expectation from SS73 is
\begin{equation}\label{eq:tau0}
    \tau_{0, \, \mathrm{SS73}} = \frac{1}{c}\Bigg(\frac{45G}{16\pi^6hc^2}\Bigg)^{1/3} \big(2700\mathrm{\AA}\big)^{4/3} \chi^{4/3} M_{\mathrm{BH}}^{1/3} \dot{M}_{\mathrm{BH}}^{1/3}.
\end{equation}
The SS73 disk-size normalization, $\tau_{0,\,\mathrm{SS73}}$, is dependent on the mass of the central black hole, \Mbh\, the accretion rate, $\Mdot = L_{\mathrm{Bol}}/\eta c^2$, where the radiative efficiency $\eta=0.1$ is assumed and $L_{\mathrm{Bol}}$ is the bolometric luminosity. The quantity $\chi$ is a geometrical factor accounting for the flux-weighted mean radius and is $\chi=2.49$. Alternately, a larger value for $\chi$ is obtained if the flux is emitted from a single annulus, $\chi$=4.97 \citep{Kammoun2021}. In this work we use the smaller $\chi$=2.49 for our main analysis; however, we note that this is one of the theoretical uncertainties of the RM disk interpertation. We adopt the normalization wavelength of 2700\AA\ based on the UVIS F275W filter pivot wavelength of 2704~\AA. We choose a normalization of 2700\AA\ to make $\lambda_{\rm opt}^{\beta} - \lambda_{\rm uv}^{\beta}$ in Equation~6 close to unity for the $r$-band (rest-frame), and thus the best-fit $\tau_0$ close to the measured rest-frame lag.

\startlongtable
\begin{deluxetable*}{cc|cccc|cccc}
\tablecaption{Significant rest-frame UV-optical Lag measurements \label{tab:table4}}
\tablehead{\colhead{} &
\multicolumn{4}{r}{\jav} &
\multicolumn{5}{c}{\cream}\\
\colhead{RMID} & \colhead{Redshift} & \colhead{$\tau_{uv-g}$} & \colhead{$\tau_{uv-r}$} & \colhead{$\tau_{uv-i}$} & \colhead{$\tau_{uv-z}$} & \colhead{$\tau_{uv-g}$} & \colhead{$\tau_{uv-r}$} & \colhead{$\tau_{uv-i}$} & \colhead{$\tau_{uv-z}$}\\
\colhead{}&\colhead{}&\colhead{days}&\colhead{days}&\colhead{days}&\colhead{days}&\colhead{days}&\colhead{days}&\colhead{days}&\colhead{days}
}
\startdata
267 & 0.588 &  ... &  $2.3_{-3.2}^{+1.9}$ &  $-2.0_{-3.3}^{+4.3}$ &  $3.7_{-1.8}^{+2.5}$ &  ... &  $3.2_{-1.6}^{+1.5}$ &  $-0.3_{-4.4}^{+9.7}$ &  $4.1_{-1.5}^{+1.3}$\\
300 & 0.646 &  $2.6_{-3.4}^{+4.1}$ &  $4.9_{-2.1}^{+2.4}$ &  ... &  $2.4_{-2.4}^{+3.0}$ &  $4.8_{-3.4}^{+3.6}$ &  $5.4_{-1.4}^{+2.0}$ &  ... &  $3.2_{-5.5}^{+4.4}$\\
399 & 0.608 &  - &  ... &  - &  - &  - & ... &  - &  -\\
551 & 0.681 &  - &  $3.5_{-3.6}^{+3.7}$ &  - &  ... &  - &  $6.4_{-4.4}^{+4.5}$ & - &  ...\\
622 & 0.572 &  - &  ... &  - &  ...  &  - &  $-0.6_{-1.0}^{+1.2}$ &  - &  ...\\
824 & 0.651 &  - &  $2.3_{-4.7}^{+4.3}$ &  - &  - &  - &  $4.9_{-3.6}^{+5.2}$ &  - &  -\\
840 & 0.244 &  ... &  $3.1_{-3.4}^{+2.7}$ &  ... &  $5.1_{-3.8}^{+3.9}$ &  ... &  $4.0_{-2.9}^{+2.1}$ &  ... &  $6.4_{-4.4}^{+4.6}$ \\
\enddata
\tablecomments{
\footnotesize
We have used two different symbols to distinguish the missing lags. We have identified those lag measurements that did not pass the lag significance criteria of Section \ref{sec:reliability} with ``..." and if the lag measurement was not available because the bandpass was not observed, we have identified it with ``-''.}
\end{deluxetable*}

We follow a Bayesian framework to fit a non-linear model using the software package \texttt{PyMc3} \citep{Salvatier2016} and determine the posterior distribution of accretion disk parameters described below. We use the SS73 accretion disk expectations as priors for the MCMC fit. We adopt the likelihood as a students' T-distribution, which has a heavier tails than a Normal distribution and so is more robust to outliers. The students' T-distribution is centered at the measured lag using either method (\jav\ or \cream) with the lag uncertainty.
We construct two chains with 20,000 draws, considering only the second half of the chain as post-burn-in draws. We explicitly check for divergences using the Gelman-Rubin statistics \citep{Gelman1992}.

\subsection{Accretion-Disk Size}\label{sec:t0}

\begin{figure*}[t]
\centering
\includegraphics[width=180mm]{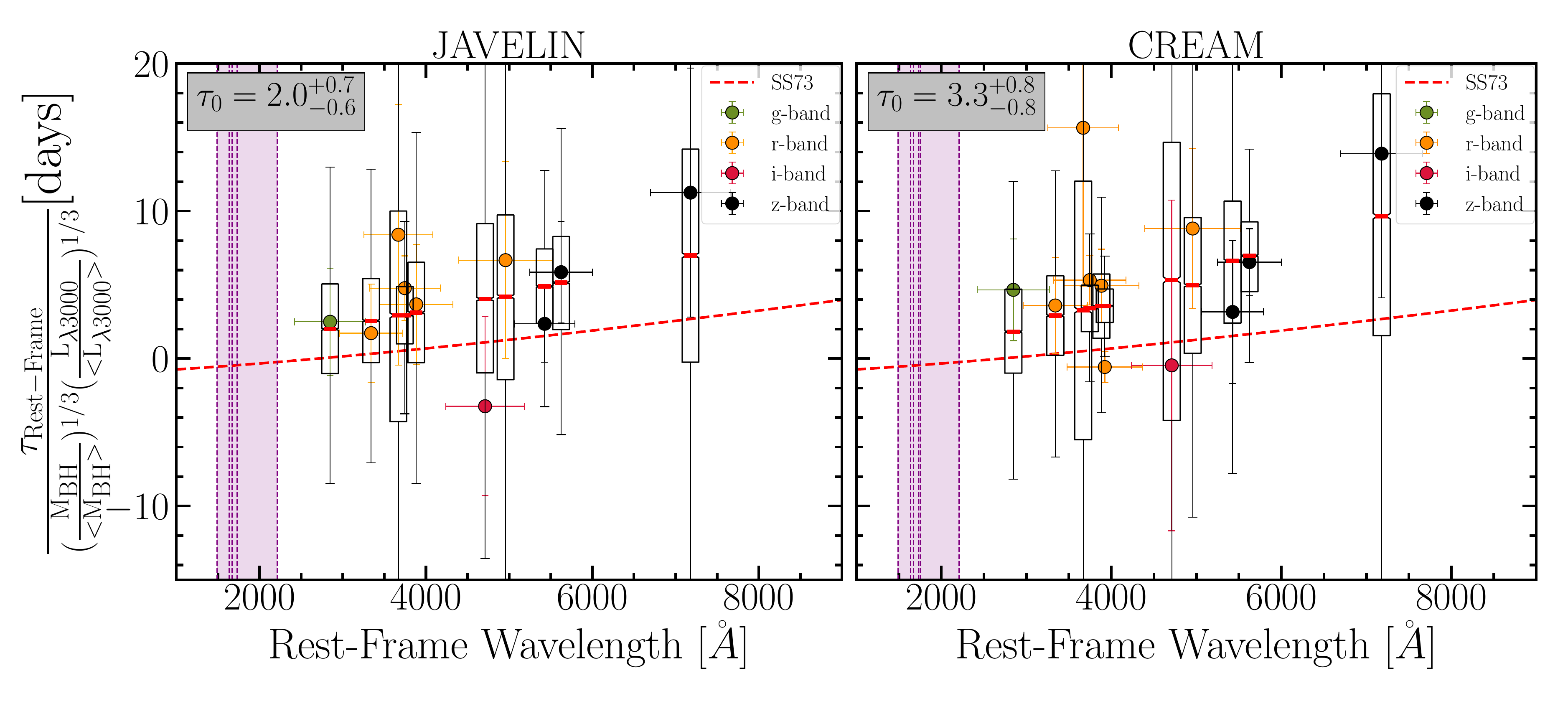}
 \caption{Rest-frame lags normalized by the ratio of \Mbh\ and \lum\ vs. wavelength using \jav\ (left) and \cream\ (right). Here we show the best-fit result (see  Section~\ref{sec:t0}), from a simple accretion-disk model where the only free parameter is the disk normalization, $\tau_0$ (see Equations~\ref{eq:tau_ss73} with fixed $\beta$=4/3). The colored points are the inter-band accretion-disk significant lags (same as figure~\ref{fig7}). The box plots illustrate the 25th and 75th percentile of the posterior predictive distribution and the thick red line marks the median of the posterior predictive distribution at each lag measurement (filled circles). The extended whiskers show the 5th to 95th percentile of the posterior predictive distribution. For the majority of our lag measurements, the model is a good representation of the data with the lag measurements overlapping within the 25th and 75th percentile of the posterior predictive distribution.}
\label{fig8}
\end{figure*}

Our first approach is to use the significant lag measurements from Section~\ref{tab:table4} and model the disk-size normalization from Equation~\ref{eq:tau_ss73} while fixing $\beta=4/3$ to match the the thin-disk value for the accretion-disk wavelength scaling, i.e., $\langle\tau\rangle\propto\lambda^{4/3}$. We use the SS73 disk normalization for mean \Mbh, $\langle\Mbh\rangle$, and mean accretion rate, $\langle\Mdot\rangle$, of the quasars with significant lags, following:
\begin{equation}\label{eq:tau0_mean}
\begin{gathered}
    \tau_{0,\,\rm SS73} = \frac{1}{c}\Bigg(\frac{45G}{16\pi^6hc^2}\Bigg)^{1/3} (2700\mathrm{\AA})^{4/3} \chi^{4/3} \Big(\frac{C_{\mathrm{Bol}}}{\eta c^2}\Big)^{1/3}  \\ \times \langle M_{\mathrm{BH}}\rangle^{1/3} \langle\lum\rangle^{1/3}
\end{gathered}
\end{equation}
where $\tau_{0,\,\rm SS73}$ is the expected value in the SS73 model, and we have assumed $\chi$=2.49, $L_{\mathrm{Bol}} = C_{\mathrm{Bol}}\,\lum$ with the bolometric luminosity correction of $C_{\mathrm{Bol}}=5.15$ from \citet{Richards2006}. We fit the disk-size normalization by combining Equations~\ref{eq:tau_ss73} and \ref{eq:tau0_mean}:
\begin{equation}\label{eq:tau_fit}
\begin{gathered}
    \tau = \tau_{0}\Bigg[\big(\frac{\lambda_{\mathrm{opt}}}{2700\mathrm{\AA}}\big)^{\beta} - \big(\frac{\mathrm{\lambda_{uv}}}{2700\mathrm{\AA}}\big)^{\beta}\Bigg]
    \\ \times \left({\frac{\Mbh}{\langle\Mbh\rangle}}\right)^{1/3}
    \left({\frac{\lum}{\langle\lum\rangle}}\right)^{1/3}
\end{gathered}
\end{equation}
Here we use the monochromatic luminosity \lum\ as a proxy for \Mdot\ and have folded the constants into the disk-size normalization, $\tau_{0}$. In this section, we refer to the best-fit disk size normalization as $\tau_0$, whereas the $\tau_{0,\,\mathrm{SS73}}$ describes the SS73 expectation for disk size normalization. The values for \Mbh\ and \lum\ are taken from Table~\ref{tab:table1}.
Figure~\ref{fig8} shows the result of the single fit and the posterior predictive distribution. We use Equation~\ref{eq:tau0_mean} to compute an accretion-disk normalization prior of 0.5$\pm$0.1 days for $\langle\rm logM_{\rm BH}\rangle=7.7 M_{\odot}$ and $\langle\lum\rangle$ = 44.3 for the \jav\ significant lags and an accretion-disk size of 0.5$\pm$0.1 days for $\langle\rm logM_{\rm BH}\rangle=7.7 M_{\odot}$ and $\langle\lum\rangle$ = 44.4 for the \cream\ significant lags sample (also see Table~\ref{tab:table1}). We use these values as a prior for the single-parameter fit, $\tau_{0,\,\rm SS73}$. Using \jav\ significant lags, we find the best-fit disk normalization from the median of the disk normalization posterior to be $2_{-0.6}^{+0.6}$~days, which is a factor of $\sim$4 larger than the mean disk normalization expectation value, and considering the uncertainties, the deviation significance level is $\sim$3 $\sigma$. Using significant \cream\ measurements, we find a slightly larger best-fit value of $3.3_{-0.8}^{+0.8}$~days, approximately 6.5 times larger than the mean SS73 disk size normalization.

\begin{figure*}[t]
\centering
\includegraphics[width=160mm]{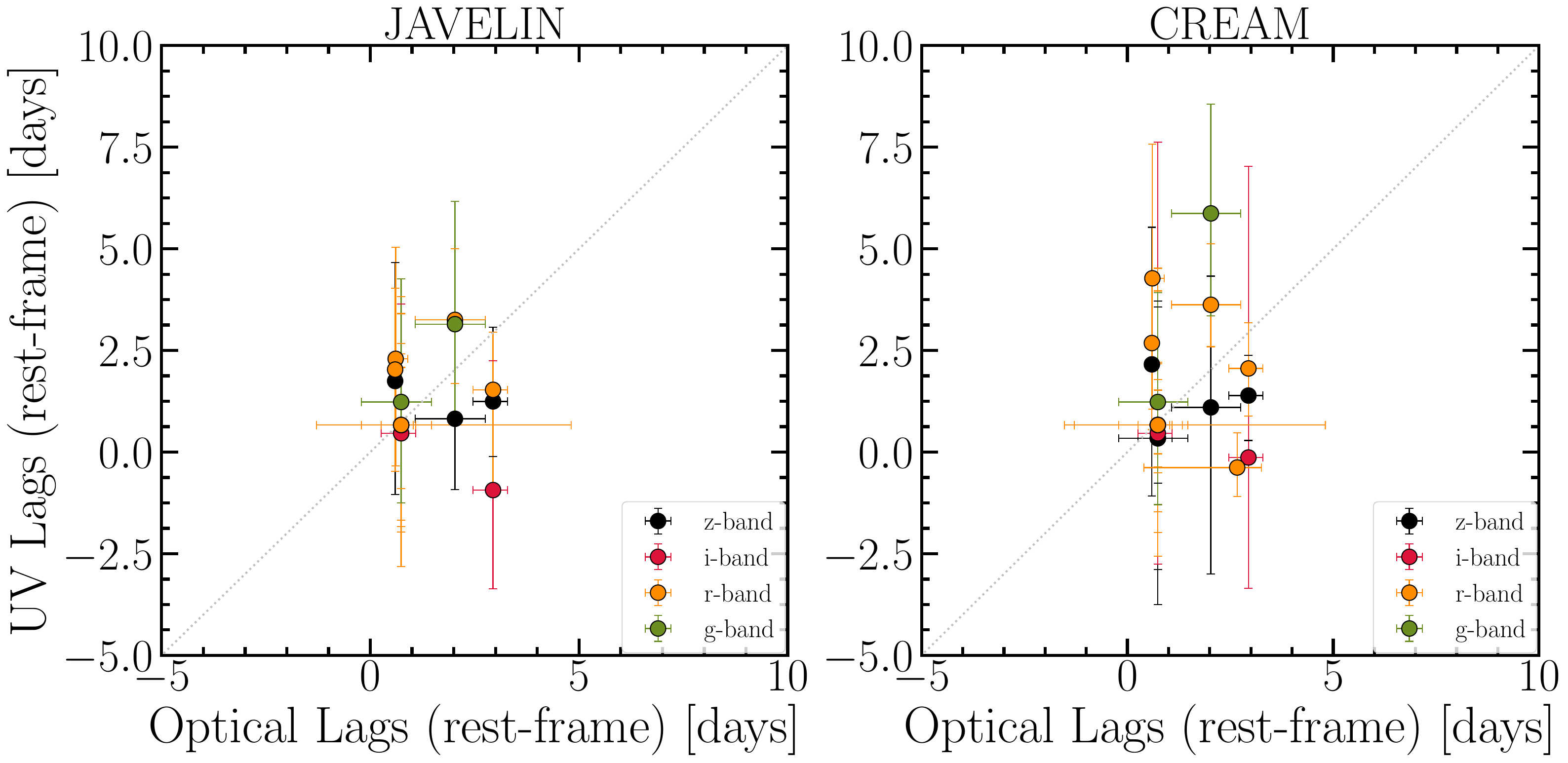}
 \caption{Rest-frame UV-optical lags vs. optical lags measured from \citet{Homayouni2019} using \jav\ (left) and \cream\ (right). For this comparison, we have converted the UV-optical from the current study as well as the optical $g$ and $i$ lags from \citet{Homayouni2019} to a common disk size, where this common disk size corresponds to the relative distance differences between 2700 \AA\ and 5100 \AA. On average our UV-optical lags are in agreement with each other. Colored points correspond to each UV-optical lag. The gray dotted line shows the 1:1 ratio.}
\label{fig8b}
\end{figure*}

Figure~\ref{fig8b} compares our UV-optical lags with optical-optical continuum lags measured for the same quasars in our previous work \citep{Homayouni2019}.
To perform a one-on-one comparison between the UV-optical lags from this study and the earlier optical lag measurements, we translated the UV-optical lag measurements to
a disk lag between continuum emission at 2700 \AA\ and 5100 \AA\, using a pivot wavelength of 2700 \AA\ and $\beta=4/3$ (see Equation~\ref{eq:tau_ss73}).
We applied the same conversion to the optical lags of \citet{Homayouni2019} measured between the $g$ and $i$ optical bands.
Our UV-optical lags are consistent with the optical lags measured for these targets (excepting one object, not shown in Figure~\ref{fig8b}, that had a negative optical-optical lag measured by \citealp{Homayouni2019}).
The observed consistency is further confirmed by our best-fit results for the temperature profile slope, $\beta$ (see Sections~\ref{sec:beta} and \ref{sec:multi_param}), since the best-fit $\beta \simeq 4/3$ implies consistent UV-optical and optical-optical lags.

High-cadence UV-optical reverberation mapping studies of local AGNs have frequently reported larger disk sizes than the standard thin-disk prediction \citep{Edelson2015, Fausnaugh2016, Cackett2018, McHardy2018, Edelson2019}. These studies report average lags that are larger by a factor of $\approx 3-4$ than SS73 predictions even after accounting for diffuse BLR contamination affecting the U-band wavelengths (see Section~\ref{sec:blr}).
On the other hand, ``industrial-scale'' photometric monitoring projects with larger and more diverse samples of quasars show average disk lags that are consistent with the SS73 model, but with significant scatter about the mean \citep{Mudd2018,Homayouni2019, Yu2020}.

At first glance, our large continuum lags agree with the previous work on nearby AGNs with disk sizes larger than the SS73 expectation. But it turns out that the quasars of this work represent only a limited subset of the larger SDSS-RM sample in terms of their measured disk sizes. The UV-optical lags are consistent with the optical-optical lags of the same quasars, as shown in Figure \ref{fig8b}, and so our measurements are consistent with a small sample that is preferentially drawn from the high side of the large scatter in disk lags among the broader quasar population. In other words, our small sample of UV-optical targets are consistent with being biased to only the high-lag portion of the broader range of quasar disk sizes. There is no obvious bias in our \hst\ sample selection that would prefer long UV-optical disk lags, and so we instead assume that this is simply a random result of selecting a small sample.

There are several possible explanations for disk lags being larger than the SS73 expectation in some subsets of quasars. \citet{Chelouche2013} argues that contribution from widespread diffuse nebular emission can increase measured continuum lags. We find evidence for this effect by diffuse nebular and iron emission, but only in specific wavelength regions, and so our work does not support the idea that diffuse nebular emission has a widespread effect on continuum lags at all wavelengths. A different reprocessing geometry, i.e. a larger $\chi$ factor in Equation~\ref{eq:tau0}, might also lead to larger continuum lags. \citet{Kammoun2021} consider reprocessing of emission from a point-source, lamp-post corona by a Novikov-Thorne general relativistic disk, including the effects of disk ionization and a potentially large height of the corona above the disk, and obtain results consistent with a larger $\chi$ factor.  Their results also do not rule out an extended corona. 
More complicated disk reprocessing, like the magnetic-coupling model of \citet{Sun2020}, would also increase the measured lag in some quasars. \citet{Li2021} show that this magnetic-coupling model is consistent with observations of the full sample of disk lags, with lower luminosity AGN typically having longer lags.
The luminosity distributions of our samples are broad enough that we cannot conclusively test this theory.  Our sample of AGN UV-optical lags have a mean of $\rm log(\lambda L_{3000}) = 44.3 \pm 0.5$, while the significant optical-optical lags from the sample's parent population in \citet{Homayouni2019} have a mean of $\rm log(\lambda L_{3000}) = 44.4\pm 0.6$.  More measurements across a wide range of luminosities are needed to test the theory of \citet{Li2021}.




\subsection{Accretion-Disk Temperature Profile}\label{sec:beta}
\begin{figure*}[t]
\includegraphics[width=180mm]{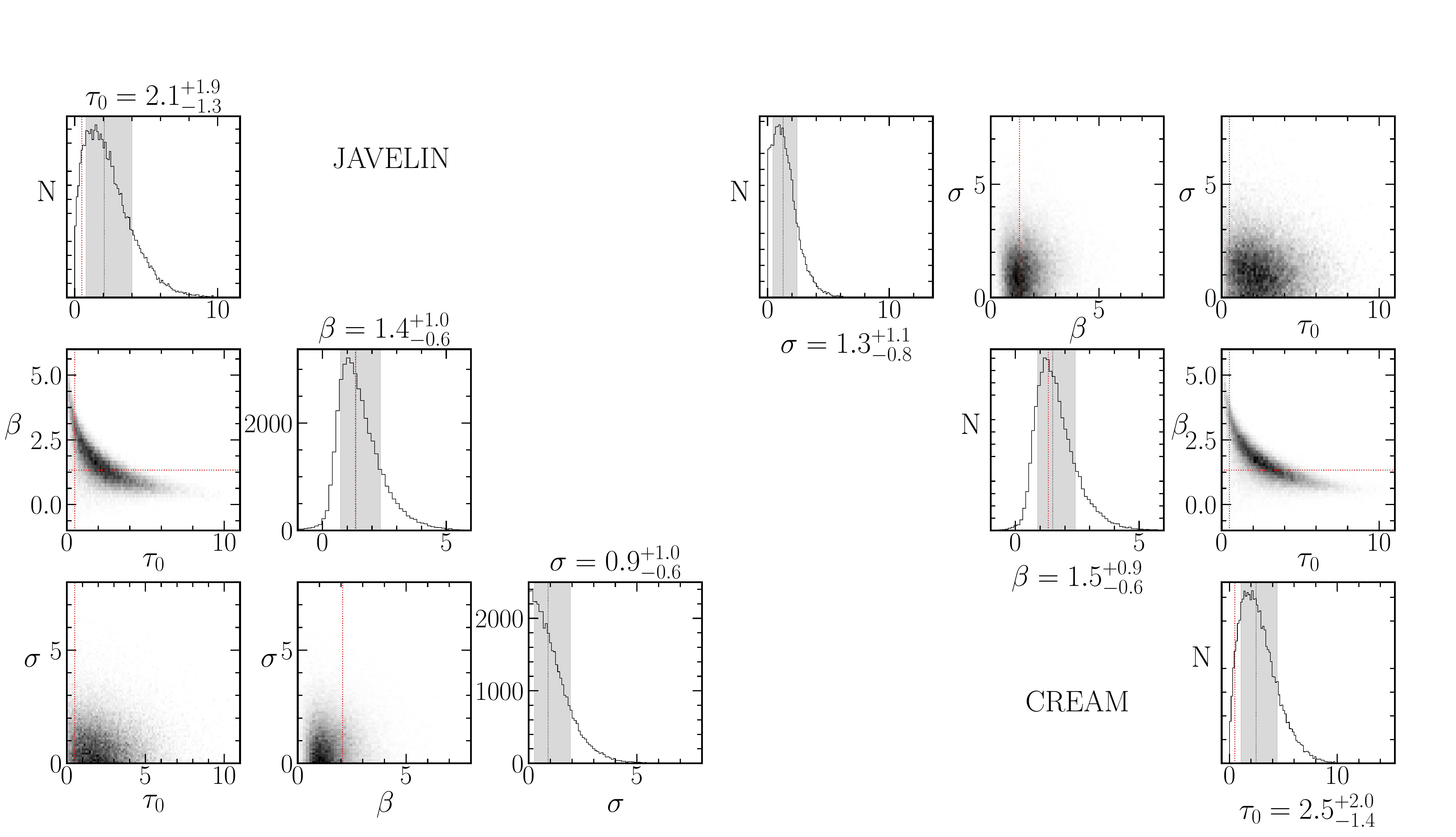}
\caption{Posterior distribution of disk-size normalization $\tau_{0}$, wavelength scaling $\beta$, and excess scatter $\sigma$ reverberation lags measured from \jav\ (left) and \cream\ (right). The red dotted lines show the prior for each parameter and the black dotted line illustrates the best-fit value computed from the median of the posteriors. The gray shaded region shows the 16th to 84th percentile for each parameter. Using our non linear regression fit, we find disk sizes that are larger than the SS73 expectation and a wavelength scaling that is consistent with $\propto{\lambda}^{4/3}$. The fit has $\sigma = 0.9_{-0.6}^{+1}$ days scatter when using \jav\ and $1.3_{-0.8}^{+1.1}$~days when using the \cream\ measurements.}
\label{fig9}
\end{figure*}

The standard thin-disk model (see Equation~\ref{eq:tau_ss73}) predicts an accretion disk structure, which can be probed by the irradiated wavelength corresponding to the measured lag, as $\tau \propto \lambda^{4/3}$. Our UV-optical lags can probe this wavelength scaling, where the measured lags target different regions of the accretion disk in the quasar rest-frame.

We fit Equation~\ref{eq:tau_fit} to the observed lags reported in Table~\ref{tab:table4}, allowing the disk size normalization, $\tau_0$, and wavelength scaling, $\beta$, to be free parameters. We adopt a bounded normal prior for the disk-fit parameters $\tau_0$ and $\beta$. Similar to Section~\ref{sec:t0}, we assume the likelihood Students' T-distribution (see Section \ref{sec:discussion}), with three degrees of freedom ($\nu = 2$) centered at the measured lag. The fit also allows for an excess scatter.
We use a half-Cauchy distribution to simultaneously fit $\sigma_{\mathrm{Excess}}$ in our non-linear regression fitting approach.

\begin{figure*}[t]
\includegraphics[width=180mm]{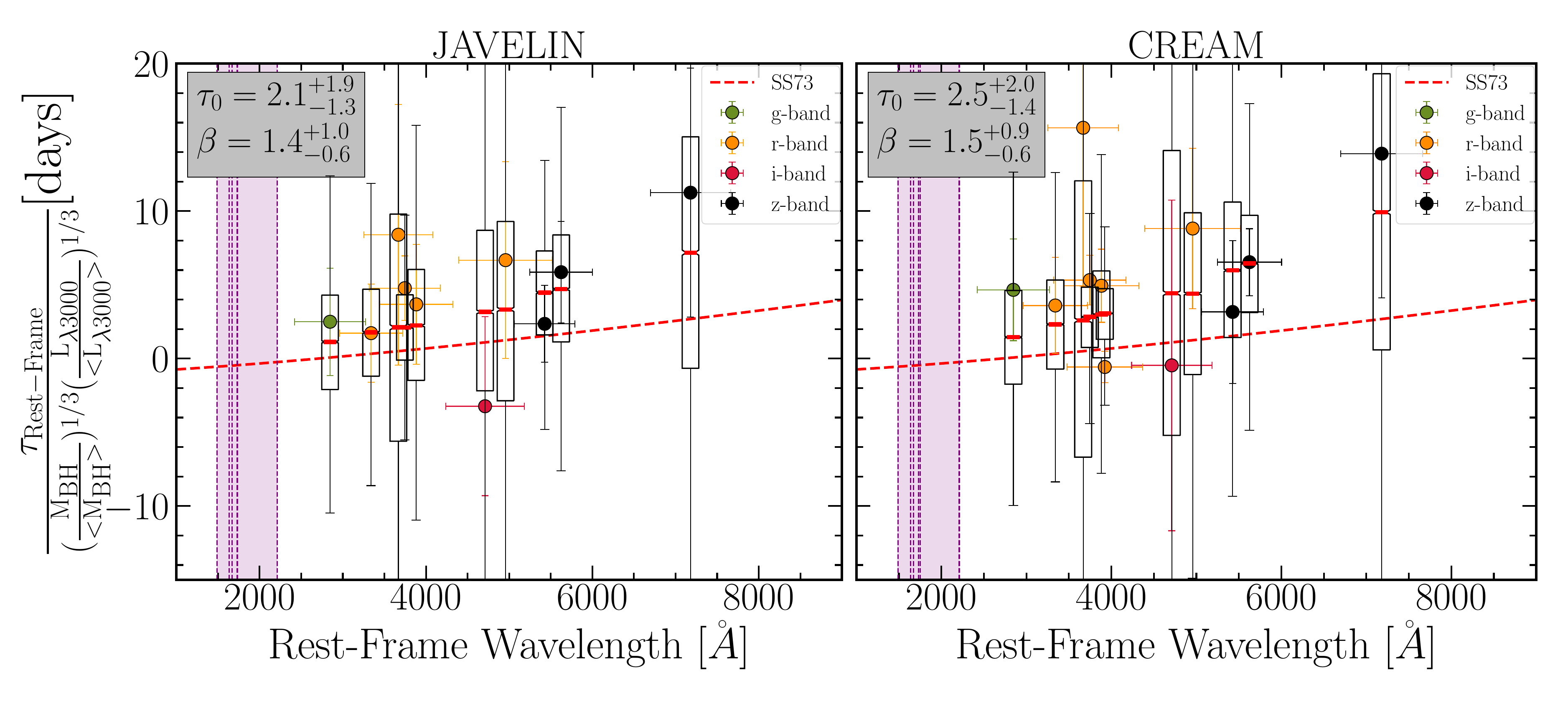}
\caption{Rest-frame lags \textbf{normalized by the ratio of \Mbh\ and \lum} versus wavelength, with a best-fit line that allows both $\tau_0$ and $\beta$ to be free parameters (analogous to the $\tau_0$-only best-fit line in Figure~\ref{fig8}), except here we have allowed both $\tau_0$ and $\beta$ to be free parameters. The red shaded region shows the SS73 expectation for the minimum and maximum of our sample's \Mbh\ and $\dot{M}_{\rm BH}$.
}
\label{fig11}
\end{figure*}

\begin{figure*}[t]
\includegraphics[width=180mm]{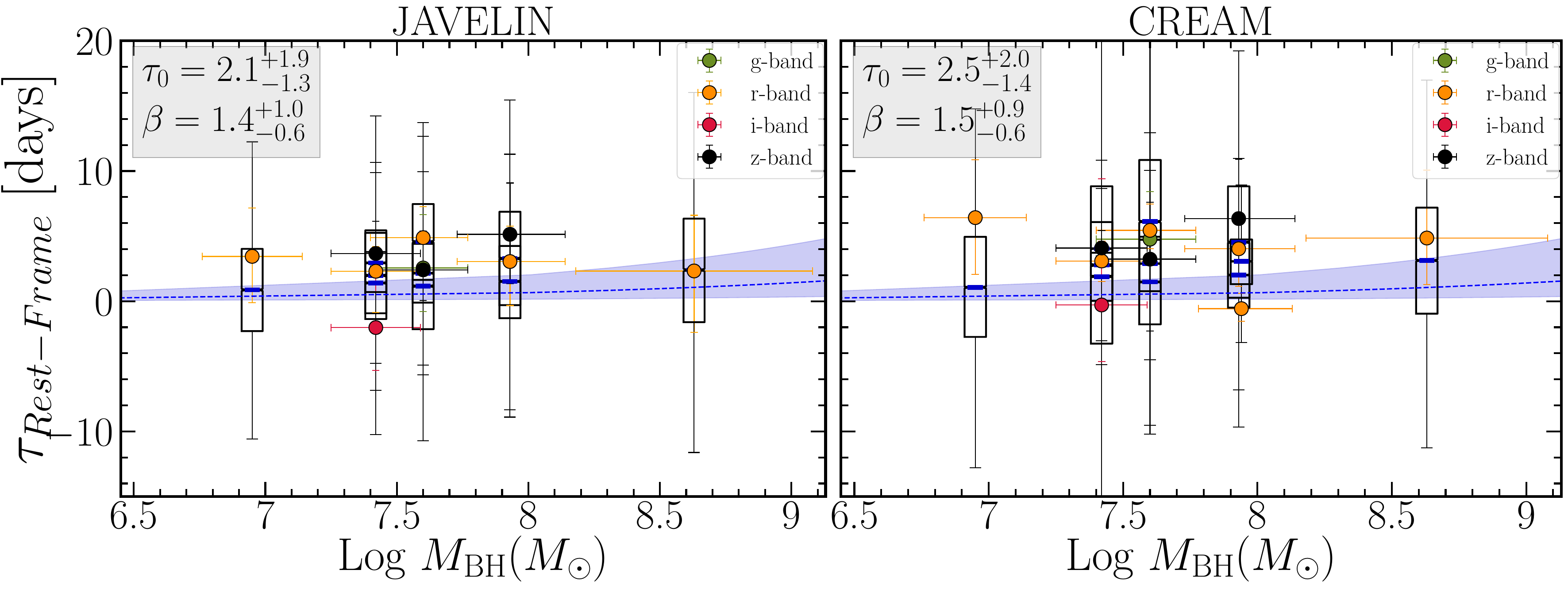}
\caption{Accretion disk lags as a function of \Mbh. Here we use the posterior predictive distribution, shown as box plots, to predict how well the data match the best-fit model. The blue-dashed line illustrates the SS73 model expectation as a function of \Mbh\ for mean monochromatic luminosity, \lum, mean inner-disk wavelength, $\lambda_{\mathrm{UV}}$, and mean outer-disk wavelength, $\lambda_{\mathrm{opt}}$. We have overlapping box plots due to multiple measurements of disk lags for the same target. Even though we have few targets to allow the \Mbh\ exponent in Equation~\ref{eq:tau_ss73} be a free parameter, we test whether the two-parameter fit shows any variation with changing \Mbh. The posterior predictive distribution reports consistent lags for different \Mbh\ measurements, consistent with both no \Mbh\ dependence and $\Mbh^{1/3}$.
}
\label{fig12}
\end{figure*}

Figures~\ref{fig9} and \ref{fig11} shows the result of our fits for both disk size and temperature profile. Using \jav\ significant lags, we find smaller best-fit values compared to the fit reported in Section~\ref{sec:t0}, though consistent within the 1$\sigma$ errorbar. The disk size, $\tau_{0} = 2.1_{-1.3}^{+1.9}$ days, is a factor of $\sim\,$4 larger than the SS73 model expectation. We find a best-fit temperature scaling of $\beta = 1.4_{-0.6}^{+1}$. As for significant \cream\ lag measurements, we find $\tau_{0} = 2.5_{-1.4}^{+2}$ days (a factor of $\sim\,$5 larger than SS73 model expectation) and $\beta = 1.5_{-0.6}^{+0.9}$. Using our two methods of lag analysis, we find the best-fit value for the wavelength scaling is consistent with the standard thin disk model approximation of $\beta=4/3$. We additionally find an excess scatter of $\sim$1 day (see Figure~\ref{fig9}), which corresponds to any unknown sources of scatter, likely related to the bolometric correction/radiative efficiency. Also, Figure~\ref{fig12} shows the posterior predictive distribution in connection to \Mbh.

Similar to Section~\ref{sec:t0}, we adopt \lum\ as a proxy for \Mdot. However, $\Mdot = L_{\mathrm Bol}/\eta c^2 = C_{\mathrm{Bol}}\lum/\eta c^2$ probably includes the largest source of uncertainty in fitting an accretion-disk model, with 0.5~dex scatter for conversion from \lum\ to \Mdot\ \citep{Richards2006, Runnoe2012}. Furthermore, the efficiency $\eta$, is commonly adopted to be 0.1 for highly accreting quasars (e.g. \citealp{Soltan1982}). However, individual quasars are likely to have a large range of efficiencies \citep{Davis2011, Sun2015}. Here, we use our two-parameter posteriors to obtain a distribution for the $C_{\mathrm{Bol}}/\eta$ ratio. We use the $\tau_0$ posterior and Equation~\ref{eq:tau0_mean} to obtain the $C_{\mathrm{Bol}}/\eta$ posterior. In general, our disk measurements are not sufficient to constrain the accretion-rate conversion parameters. Our result is broadly consistent (within 1$\sigma$) with the empirical value of $C_{\mathrm{Bol}}$ = 5.15 and $\eta$ = 10\%, but the posterior distribution has a long tail that is not particularly constraining on the allowed $C_{\mathrm{Bol}}/\eta$.

\subsection{Disk Size Dependence on \Mbh\ and \lum}\label{sec:multi_param}

To go beyond a fit that is limited to disk-size normalization and wavelength scaling, we perform a nonlinear regression to fit for the relationship between the disk size, \Mbh, and \lum. We examine if the measured continuum lag depends on \Mbh\ and \Mdot, as expected for the SS73 model, by fitting a non-linear MCMC regression in three different and independent steps. First we test for connections to \Mbh\ by fixing $\dot{M}^{1/3}$ (i.e., $\lum^{1/3}$) and fit Equation~\ref{eq:tau_fit} for $\tau_{0,\,\rm SS73}$ and $\beta$ while allowing the \Mbh\ exponent be a free parameter, $\Mbh^{\gamma}$:
\begin{equation}\label{eq:tau_fit_mbh}
\begin{gathered}
    \tau = \tau_{0}\Bigg[\big(\frac{\lambda_{\mathrm{opt}}}{2700\mathrm{\AA}}\big)^{\beta} - \big(\frac{\mathrm{\lambda_{uv}}}{2700\mathrm{\AA}}\big)^{\beta}\Bigg]
    \\ \times \left({\frac{\Mbh}{\langle\Mbh\rangle}}\right)^{\gamma}
    \left({\frac{\lum}{\langle\lum\rangle}}\right)^{1/3}
\end{gathered}
\end{equation}
Second, we fit for a dependence of disk lags on $\tau_{0,\,\rm SS73}$, $\beta$, and the accretion rate  while fixing the \Mbh\ exponent to the SS73 expectation, $\Mbh^{1/3}$ and allow the dependence of the disk size with observable luminosity, \lum, be a free parameter $\lum^{\delta}$ as:

\begin{equation}\label{eq:tau_fit_l3000}
\begin{gathered}
    \tau = \tau_{0}\Bigg[\big(\frac{\lambda_{\mathrm{opt}}}{2700\mathrm{\AA}}\big)^{\beta} - \big(\frac{\mathrm{\lambda_{uv}}}{2700\mathrm{\AA}}\big)^{\beta}\Bigg]
    \\ \times \left({\frac{\Mbh}{\langle\Mbh\rangle}}\right)^{1/3}
    \left({\frac{\lum}{\langle\lum\rangle}}\right)^{\delta}
\end{gathered}
\end{equation}

Finally, in the third step of our fit we allow both the \Mbh\ and \lum\ exponents be free parameters (i.e., $\tau\propto\Mbh^{\gamma}\lum^{\delta})$ with fixed $\beta = 4/3$ as expected from SS73 and from our two-parameter fit in Section~\ref{sec:beta}. We follow an independent and step-by-step approach to fitting to build better intuition and avoid overinterpreting multiparameter fits with large uncertainties, given our small sample size and the large uncertainties associated with $\eta$ and $C_{bol}$.
\startlongtable
\begin{deluxetable*}{cccccc}
\tablecaption{Fits to black hole mass and \lum \label{tab:table5}}
\tablehead{
\colhead{Free Parameters} & \colhead{$\tau_0$} & \colhead{$\beta$} & \colhead{$\gamma$} & \colhead{$\delta$} & \colhead{$\sigma$}
}
\startdata
$\tau_0$, $\beta$, $\gamma$ & $1.7_{-1.2}^{+1.9}$ & $1.3_{-0.7}^{+1.1}$ & $0.3_{-0.8}^{+0.9}$& fixed (1/3) & $0.9_{-0.7}^{+1.1}$\\
$\tau_0$, $\beta$, $\delta$ & $1.8_{-1.2}^{+1.9}$ & $1.1_{-0.6}^{+1.0}$ & fixed (1/3) & $0.3_{-0.5}^{+1.0}$ & $1_{-0.7}^{+1.1}$ \\
$\tau_0$, $\gamma$, $\delta$ & $1.3_{-1}^{+2.0}$ & fixed(4/3) & $0.3_{-1.0}^{+1.5}$ & $0.6_{-0.8}^{+1.6}$ & $1.1_{-0.7}^{+1.2}$\\
\enddata
\end{deluxetable*}
Our set of black-hole masses is obtained from RM \Hb\ masses \citep{Grier2017} and only one of our targets, RM824, has its black-hole mass measured using the single-epoch method \citep{Shen2019a}. To perform the nonlinear MCMC regression for the three-parameter fit, with $\tau_{0,\,\rm SS73}$, $\beta$, and $\gamma$, we provide the \Mbh\ prior as a normal distribution centered at the measured \Mbh\ with uncertainties as the width from Table~\ref{tab:table1}. To perform the three-parameter fit that includes $\tau_{0,\,\rm SS73}$, $\beta$, and $\delta$, we incorporate only the \lum\ measurements and uncertainties as a normal distribution prior. Similar to previous discussions in Sections~\ref{sec:t0} and \ref{sec:beta}, we report an excess scatter, $\sigma$, for the fit. Table~\ref{tab:table5} provides a brief summary of these different fitting approaches using the \jav\ lags results. In general, we find the best-fit values are consistent with theoretical SS73 expectations, but with large uncertainties that are similarly consistent with a wide range of relationships between disk size, black-hole mass, and accretion rate. The three-parameter fit involving both \Mbh\ and \lum\ shows the highest scatter.

\section{Summary}

We have presented results from an intensive UV-optical photometric monitoring campaign of eight SDSS-RM quasars. The selected sample has the advantages of a wide range of Eddington ratio, reliable black hole mass from the first-year of the SDSS-RM monitoring program \citep{Grier2017}. Our study of UV-optical disk measurement is the first study to go beyond $z>0.3$. Our set of UV lightcurves have an every-other-day (2-day) cadence from \textit{HST} UVIS F275W and coordinated ground-based monitoring for up to four optical bands over three months of monitoring. We use these sets of photometric lightcurves to measure UV-optical continuum lags and to study the accretion disk structure and its connection to accretion rate. We report UV-optical lag results from two lag-identification methods, \jav\ and \cream. We use statistical criteria to ensure that we select significant lags that are arising from physical reverberation. Our main results are as follows:

\begin{enumerate}

    \item Significant continuum lags are detected between the UV at $\lambda\,$2704 \AA, and optical broad-band $g,r,i$ and $z$ filters at 4686, 6166, 7480, 8932 \AA. Due to lag-significance criteria, not all four inter-band lags were found to be significant measurements for every target (with some limitations due to observation design). In general, the time delay observation is found to be consistent with a disk-stratification model where $\tau_{uv-g} < \tau_{uv-r} < \tau_{uv-i} < \tau_{uv-z}$.

    \item We find an excess of large lags (rest-frame lags~$>+10$~days and $<-10$~days) that overlap with the diffuse Balmer continuum window at $\lambda\,$ 3500-3900 \AA\ and the diffuse iron continuum windows at $\lambda\,$ 4434-4684 \AA\ and $\lambda\,$ 5100-5477 \AA. These outlier lags are a factor of $\approx$ 2.5 larger than the mean \jav\ significant lags of 2.8~days and a factor of $\approx$ 3.8 times larger than mean \cream\ significant lags of 3.8~days. We additionally have one source with a long lag that is associated with significant contamination from the \Ha\ emission line.

    \item The best-fit UV-optical disk-size normalization is found to be consistently larger than the SS73 theoretical expectation in all the three fitting approaches. From the simple one-parameter fit, we found disk sizes that are $\sim$4-6 times larger than SS73 expectation of 0.5~days. Using the two-parameter fitting approach, we found disk normalizations that are $\sim$4-5 times larger, and finally from the three-parameter fits, we found disk-size normalizations that are $\sim$2-3 times larger than the standard thin disk model, assuming $\chi$=2.49. However, larger disks can also be explained by larger $\chi=4.97$ for a single flux annulus, and could reduce these differences by half.

    \item We show that our UV-optical lags are consistent with the optical-optical lags as measured previously for the same quasars \citep{Homayouni2019}. Our quasars are selected from a broad diversity of the SS73 disk sizes, and these measurements are consistent with being \textbf{drawn} from the high-lag portion of the SDSS-RM sample.

    \item The trend of increasing lag as a function of wavelength is consistent with the standard thin-disk expectation of $\tau\propto\lambda^{4/3}$. We found a best-fit value for the wavelength scaling $\beta = 1.4_{-0.6}^{+1}$ using the \jav\ method and a slightly larger, but consistent $\beta = 1.5_{-0.6}^{+0.9}$ using \cream\ measurements.

    \item Assuming that continuum lags scale with black-hole mass as $\tau\propto\tau_0\Mbh^{\gamma}$, $\tau\propto\tau_0\lum^{\delta}$ and $\tau\propto\tau_0\Mbh^{\gamma}\lum^{\delta}$,  we examined the dependency upon \Mbh\ and \lum\ from three different fitting approaches. We found that the disk size is connected to \Mbh\ consistent with the SS73 expectation (i.e., power-law slope of 1/3). We found the dependence to \lum\ is also consistent with the theoretical value from SS73; however, the best-fit values for mass and luminosity dependence have higher uncertainty and excess scatter when they are simultaneously allowed to be a free parameter in the fit.

\end{enumerate}

Our new measurements represent a new advance in ``industrial-scale" multi object UV-optical accretion-disk size measurements from \hst\ observations. Our measured disk sizes are broadly consistent with the SS73 disk model.  We demonstrate that fitting only the disk normalization results in larger disks by a factor of $\sim$5-6 while fitting a comprehensive accretion disk including the color profile and mass and luminosity results in disks that are $\sim$2 times larger, although with larger uncertainties. This motivates future work to better measure bolometric luminosity and radiative efficiency alongside accretion disk sizes and black hole mass.

\software{CREAM \citep{Starkey2016}, Javelin \citep{Zu2011}, PyceCREAM (https://github.com/dstar-key23/pycecream), PyMc3 (https://docs.pymc.io/notebooks/-GLM-robust.html)}

YH, JRT, and GFA acknowledge support from NASA grants HST-GO-15650 and 18-2ADAP18-0177 and NSF grant CAREER-1945546. KH acknowledges support from STFC grant ST/R000824/1. CJG acknowledges support from NSF grant AST-2009949. YS acknowledges support from NSF grants AST-1715579 and AST-2009947.  PH acknowledges support from the Natural Sciences and Engineering Research Council of Canada (NSERC), funding reference number 2017-05983. LCH was supported by the National Science Foundation of China (11721303, 11991052) and the National Key R\&D Program of China (2016YFA0400702).

\appendix
Our coordinated ground-based observations (see Section~\ref{sec:ground}) have flux measurements with a median signal-to-noise ratio (SNR) of $\sim$ 16, where a typical variation of SNR is $\sim$ 3 for our set of lightcurves. 
Here we assess the effects of the lightcurve SNR on lag recovery rate using simulated lightcurves.

To generate our synthetic optical lightcurves, we start from the \jav\ DRW modeled UV lightcurve for our sample of significant \jav\ lags in Table~\ref{tab:table4}. For each optical lightcurve simulation, we assign typical noise to the DRW model, where the noise is drawn from a random normal distribution with the median and NAMD of the observed lightcurves SNR. At each epoch, we resample the flux using a Gaussian normal distribution with the model mean flux and a dispersion equal to the flux uncertainty determined by the noise. We scale the synthetic lightcurve variance to match the RMS variability of the observed lightcurves. To mimic the UV-optical lag, we shift the optical lightcurves by representative accretion disk lags of 1, 2, 4, 8, and 16 days. To realistically model the responding optical lightcurves with a broader disk response, we convolve these optical lightcurves with a Gaussian kernel with widths that are 20\% of the input lags. This accounts for the wavelength-dependent aspect of the transfer function as demonstrated by \citet{Starkey2016} where the longer wavelength response has a broader transfer function. Finally, the synthetic lightcurves are down-sampled to have similar cadence as the observed cadence reported in Table~\ref{tab:table2}. To incorporate the effects of the non-uniform noise due to lunation, we down-sample the simulated lightcurves by selecting only the epochs that match the observed epochs. We then add the actual observed flux uncertainty to each simulated data point to capture similar flux uncertainties as was observed for our optical lightcurves. We simulate N=10 times per target and bandpass, totalling 500 simulated optical lightcurves.

Similarly, we generate the simulated UV lightcurves from the DRW models, resampling the flux using a random normal distribution with the model flux and a dispersion equal to square root of the sum of representative UV-lightcurve noise and the model uncertainty squared.

Finally we down-sample the UV lightcurve using the total number of subexposures in the \hst\ observations (see Table~\ref{tab:table2}).

We then use \jav\ to compute the UV-optical lag between each pair of UV and optical lightcurves, with damping time scale and transfer function width as described in Section~\ref{sec:method}. After we compute all lags for the simulated lightcurves, we identify the significant lags using the lag reliability criteria discussed in Section~\ref{sec:reliability}.

\begin{figure}[t]
\includegraphics[width=180mm]{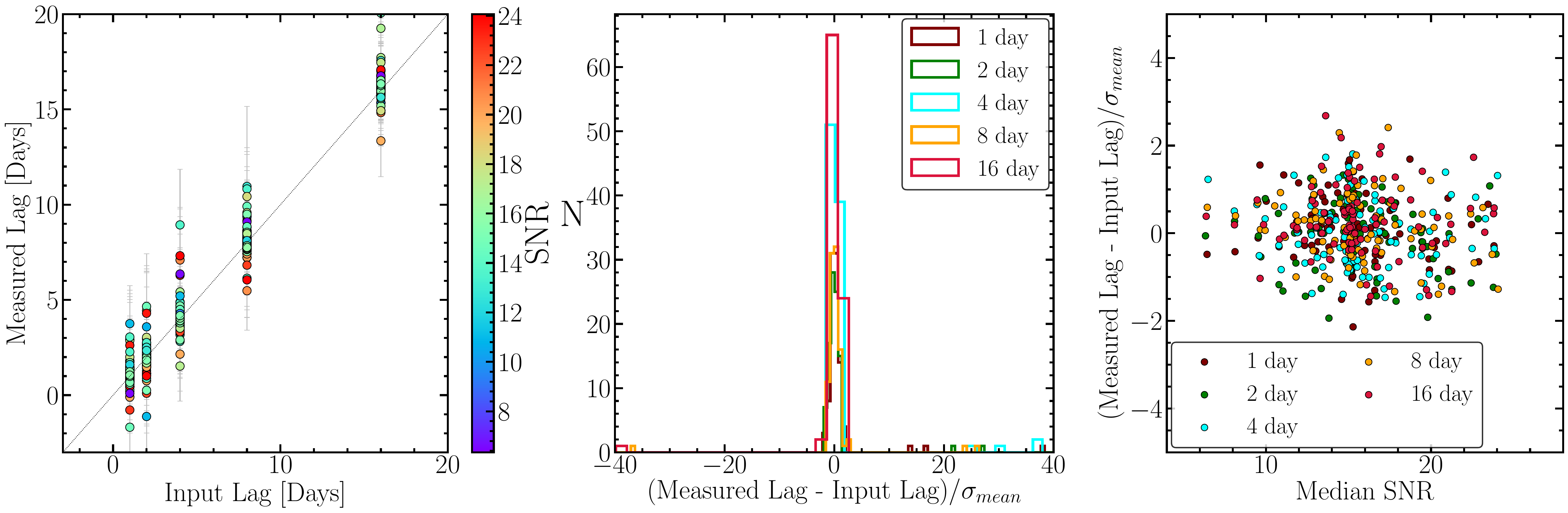}
\caption{\textit{Left:} The comparison between the measured and input lags from 500 simulated lightcurves, color-coded by the median SNR of the lightcurve fluxes. The synthetic lightcurves have added noise from our optical photometry observations. We test different input lags similar to expected accretion disk lag values. Overall we find that our measured lags are consistent with the median of each distribution overlapping with 1:1 line. \textit{Middle:} The distribution of lag difference divided by the \jav\ uncertainties illustrated by a different color for each of the input lags. The breakdown for each individual lag shows that we find that the measured lag is consistent within $<1\sigma$ for 66\%-68\% of the simulations. We also find that the measured and input lags are consistent within $<2\sigma$ for 92\% - 96\% of the simulations. Thus, we find there is no bias in the measured lag and the estimated lag uncertainties are accurate. \textit{Right:} Distribution of the lag difference weighted by lag uncertainties as a function of median lightcurve SNR. Here we use similar color-coding as the middle panel to illustrate the different input lags. For a clearer display of the results, the plot limits exclude the 4\% of cases where the lag differences are discrepant by more than 5$\sigma$. }
\label{fig13}
\end{figure}

In the end, we measured 478 significant lags from 500 synthetic UV-optical lightcurve set in Table~\ref{tab:table4}.  Figure~\ref{fig13} compares the input and recovered lags for the significant lag measurements. We find that the synthetic light curves have lags that are statistically consistent with the input lags.
Our simulations reveal that the measured and input lags are consistent for 66\%-68\% within $<1\sigma$ and also similarly consistent within $<2\sigma$ for 92\% - 96\% of the simulations.
We find no bias in lag measurement caused by the SNR in the optical lightcurves.
We also conclude that the estimated lag uncertainties are reliable since they accurately describe the differences between the input and measured lags.

\bibliography{main.bib}

\begin{thebibliography}{}
\expandafter\ifx\csname natexlab\endcsname\relax\def\natexlab#1{#1}\fi
\providecommand{\url}[1]{\href{#1}{#1}}

\bibitem[{{Bentz} \& {Katz}(2015)}]{Bentz2015}
{Bentz}, M.~C., \& {Katz}, S. 2015, Publications of the Astronomical Society of
  the Pacific, 127, 67

\bibitem[{{Blandford} \& {McKee}(1982)}]{Blandford1982}
{Blandford}, R.~D., \& {McKee}, C.~F. 1982, \apj, 255, 419

\bibitem[{{Boroson} \& {Green}(1992)}]{Boroson1992}
{Boroson}, T.~A., \& {Green}, R.~F. 1992, \apjs, 80, 109

\bibitem[{{Bradley} {et~al.}(2017){Bradley}, {Sipocz}, {Robitaille},
  {Vin{\'\i}cius}, {Tollerud}, {Deil}, {Barbary}, {G{\"u}nther}, {Cara},
  {Busko}, {Droettboom}, {Bostroem}, {Bray}, {Andersen Bratholm}, {Pickering},
  {Craig}, {Barentsen}, {Pascual}, {Conseil}, {adonath}, {Greco}, {Kerzendorf},
  {de Val-Borro}, {StuartLittlefair}, {Ogaz}, {Lim}, {Ferreira}, {D'Eugenio},
  \& {Weaver}}]{Bradley2017}
{Bradley}, L., {Sipocz}, B., {Robitaille}, T., {et~al.} 2017,
  {Astropy/Photutils: V0.4}, vv0.4,  Zenodo, doi:10.5281/zenodo.1039309

\bibitem[{{Bramich}(2008)}]{Bramich2008}
{Bramich}, D.~M. 2008, \mnras, 386, L77

\bibitem[{{Bruhweiler} \& {Verner}(2008)}]{Bruhweiler2008}
{Bruhweiler}, F., \& {Verner}, E. 2008, \apj, 675, 83

\bibitem[{{Cackett} {et~al.}(2018){Cackett}, {Chiang}, {McHardy}, {Edelson},
  {Goad}, {Horne}, \& {Korista}}]{Cackett2018}
{Cackett}, E.~M., {Chiang}, C.-Y., {McHardy}, I., {et~al.} 2018, \apj, 857, 53

\bibitem[{{Cackett} {et~al.}(2007){Cackett}, {Horne}, \&
  {Winkler}}]{Cackett2007}
{Cackett}, E.~M., {Horne}, K., \& {Winkler}, H. 2007, \mnras, 380, 669

\bibitem[{{Cackett} {et~al.}(2020){Cackett}, {Gelbord}, {Li}, {Horne}, {Wang},
  {Barth}, {Bai}, {Bian}, {Carroll}, {Du}, {Edelson}, {Goad}, {Ho}, {Hu},
  {Khatu}, {Luo}, {Miller}, \& {Yuan}}]{Cackett2020}
{Cackett}, E.~M., {Gelbord}, J., {Li}, Y.-R., {et~al.} 2020, \apj, 896, 1

\bibitem[{{Chan} {et~al.}(2020){Chan}, {Millon}, {Bonvin}, \&
  {Courbin}}]{Chan2020}
{Chan}, J.~H.~H., {Millon}, M., {Bonvin}, V., \& {Courbin}, F. 2020, \aap, 636,
  A52

\bibitem[{{Chelouche}(2013)}]{Chelouche2013}
{Chelouche}, D. 2013, \apj, 772, 9

\bibitem[{{Chelouche} \& {Zucker}(2013)}]{CheloucheZucker2013}
{Chelouche}, D., \& {Zucker}, S. 2013, \apj, 769, 124

\bibitem[{{Collier} {et~al.}(2001){Collier}, {Crenshaw}, {Peterson}, {Brandt},
  {Clavel}, {Edelson}, {George}, {Horne}, {Kriss}, {Mathur}, {Netzer},
  {O'Brien}, {Pogge}, {Pounds}, {Romano}, {Shemmer}, {Turner}, \&
  {Wamsteker}}]{Collier2001}
{Collier}, S., {Crenshaw}, D.~M., {Peterson}, B.~M., {et~al.} 2001, \apj, 561,
  146

\bibitem[{{Collier} {et~al.}(1998){Collier}, {Horne}, {Kaspi}, {Netzer},
  {Peterson}, {Wanders}, {Alexander}, {Bertram}, {Comastri}, {Gaskell},
  {Malkov}, {Maoz}, {Mignoli}, {Pogge}, {Pronik}, {Sergeev}, {Snedden},
  {Stirpe}, {Bochkarev}, {Burenkov}, {Shapovalova}, \& {Wagner}}]{Collier1998}
{Collier}, S.~J., {Horne}, K., {Kaspi}, S., {et~al.} 1998, \apj, 500, 162

\bibitem[{{Davis} \& {Laor}(2011)}]{Davis2011}
{Davis}, S.~W., \& {Laor}, A. 2011, \apj, 728, 98

\bibitem[{{Edelson} {et~al.}(2015){Edelson}, {Gelbord}, {Horne}, {McHardy},
  {Peterson}, {Ar{\'e}valo}, {Breeveld}, {De Rosa}, {Evans}, {Goad}, {Kriss},
  {Brandt}, {Gehrels}, {Grupe}, {Kennea}, {Kochanek}, {Nousek}, {Papadakis},
  {Siegel}, {Starkey}, {Uttley}, {Vaughan}, {Young}, {Barth}, {Bentz},
  {Brewer}, {Crenshaw}, {Dalla Bont{\`a}}, {De Lorenzo-C{\'a}ceres}, {Denney},
  {Dietrich}, {Ely}, {Fausnaugh}, {Grier}, {Hall}, {Kaastra}, {Kelly},
  {Korista}, {Lira}, {Mathur}, {Netzer}, {Pancoast}, {Pei}, {Pogge},
  {Schimoia}, {Treu}, {Vestergaard}, {Villforth}, {Yan}, \& {Zu}}]{Edelson2015}
{Edelson}, R., {Gelbord}, J.~M., {Horne}, K., {et~al.} 2015, \apj, 806, 129

\bibitem[{{Edelson} {et~al.}(2017){Edelson}, {Gelbord}, {Cackett}, {Connolly},
  {Done}, {Fausnaugh}, {Gardner}, {Gehrels}, {Goad}, {Horne}, {McHardy},
  {Peterson}, {Vaughan}, {Vestergaard}, {Breeveld}, {Barth}, {Bentz},
  {Bottorff}, {Brandt}, {Crawford}, {Dalla Bont{\`a}}, {Emmanoulopoulos},
  {Evans}, {Figuera Jaimes}, {Filippenko}, {Ferland}, {Grupe}, {Joner},
  {Kennea}, {Korista}, {Krimm}, {Kriss}, {Leonard}, {Mathur}, {Netzer},
  {Nousek}, {Page}, {Romero-Colmenero}, {Siegel}, {Starkey}, {Treu}, {Vogler},
  {Winkler}, \& {Zheng}}]{Edelson2017}
{Edelson}, R., {Gelbord}, J., {Cackett}, E., {et~al.} 2017, \apj, 840, 41

\bibitem[{{Edelson} {et~al.}(2019){Edelson}, {Gelbord}, {Cackett}, {Peterson},
  {Horne}, {Barth}, {Starkey}, {Bentz}, {Brandt}, {Goad}, {Joner}, {Korista},
  {Netzer}, {Page}, {Uttley}, {Vaughan}, {Breeveld}, {Cenko}, {Done}, {Evans},
  {Fausnaugh}, {Ferland}, {Gonzalez-Buitrago}, {Gropp}, {Grupe}, {Kaastra},
  {Kennea}, {Kriss}, {Mathur}, {Mehdipour}, {Mudd}, {Nousek}, {Schmidt},
  {Vestergaard}, \& {Villforth}}]{Edelson2019}
---. 2019, \apj, 870, 123

\bibitem[{{Fausnaugh} {et~al.}(2016){Fausnaugh}, {Denney}, {Barth}, {Bentz},
  {Bottorff}, {Carini}, {Croxall}, {De Rosa}, {Goad}, {Horne}, {Joner},
  {Kaspi}, {Kim}, {Klimanov}, {Kochanek}, {Leonard}, {Netzer}, {Peterson},
  {Schn{\"u}lle}, {Sergeev}, {Vestergaard}, {Zheng}, {Zu}, {Anderson},
  {Ar{\'e}valo}, {Bazhaw}, {Borman}, {Boroson}, {Brandt}, {Breeveld}, {Brewer},
  {Cackett}, {Crenshaw}, {Dalla Bont{\`a}}, {De Lorenzo-C{\'a}ceres},
  {Dietrich}, {Edelson}, {Efimova}, {Ely}, {Evans}, {Filippenko}, {Flatland},
  {Gehrels}, {Geier}, {Gelbord}, {Gonzalez}, {Gorjian}, {Grier}, {Grupe},
  {Hall}, {Hicks}, {Horenstein}, {Hutchison}, {Im}, {Jensen}, {Jones},
  {Kaastra}, {Kelly}, {Kennea}, {Kim}, {Korista}, {Kriss}, {Lee}, {Lira},
  {MacInnis}, {Manne-Nicholas}, {Mathur}, {McHardy}, {Montouri}, {Musso},
  {Nazarov}, {Norris}, {Nousek}, {Okhmat}, {Pancoast}, {Papadakis}, {Parks},
  {Pei}, {Pogge}, {Pott}, {Rafter}, {Rix}, {Saylor}, {Schimoia}, {Siegel},
  {Spencer}, {Starkey}, {Sung}, {Teems}, {Treu}, {Turner}, {Uttley},
  {Villforth}, {Weiss}, {Woo}, {Yan}, \& {Young}}]{Fausnaugh2016}
{Fausnaugh}, M.~M., {Denney}, K.~D., {Barth}, A.~J., {et~al.} 2016, \apj, 821,
  56

\bibitem[{{Fausnaugh} {et~al.}(2018){Fausnaugh}, {Starkey}, {Horne},
  {Kochanek}, {Peterson}, {Bentz}, {Denney}, {Grier}, {Grupe}, {Pogge}, {De
  Rosa}, {Adams}, {Barth}, {Beatty}, {Bhattacharjee}, {Borman}, {Boroson},
  {Bottorff}, {Brown}, {Brown}, {Brotherton}, {Coker}, {Crawford}, {Croxall},
  {Eftekharzadeh}, {Eracleous}, {Joner}, {Henderson}, {Holoien}, {Hutchison},
  {Kaspi}, {Kim}, {King}, {Li}, {Lochhaas}, {Ma}, {MacInnis}, {Manne-Nicholas},
  {Mason}, {Montuori}, {Mosquera}, {Mudd}, {Musso}, {Nazarov}, {Nguyen},
  {Okhmat}, {Onken}, {Ou- Yang}, {Pancoast}, {Pei}, {Penny}, {Poleski},
  {Rafter}, {Romero- Colmenero}, {Runnoe}, {Sand}, {Schimoia}, {Sergeev},
  {Shappee}, {Simonian}, {Somers}, {Spencer}, {Stevens}, {Tayar}, {Treu},
  {Valenti}, {Van Saders}, {Villanueva}, {Villforth}, {Weiss}, {Winkler}, \&
  {Zhu}}]{Fausnaugh2018}
{Fausnaugh}, M.~M., {Starkey}, D.~A., {Horne}, K., {et~al.} 2018, \apj, 854,
  107

\bibitem[{{Galeev} {et~al.}(1979){Galeev}, {Rosner}, \& {Vaiana}}]{Galeev1979}
{Galeev}, A.~A., {Rosner}, R., \& {Vaiana}, G.~S. 1979, \apj, 229, 318

\bibitem[{{Gelman} \& {Rubin}(1992)}]{Gelman1992}
{Gelman}, A., \& {Rubin}, D.~B. 1992, Statistical Science, 7, 457

\bibitem[{{Gonzaga}(2012)}]{Gonzaga2012}
{Gonzaga}, S. 2012, {The DrizzlePac Handbook}

\bibitem[{{Grier} {et~al.}(2017){Grier}, {Trump}, {Shen}, {Horne}, {Kinemuchi},
  {McGreer}, {Starkey}, {Brandt}, {Hall}, {Kochanek}, {Chen}, {Denney},
  {Greene}, {Ho}, {Homayouni}, {I-Hsiu Li}, {Pei}, {Peterson}, {Petitjean},
  {Schneider}, {Sun}, {AlSayyad}, {Bizyaev}, {Brinkmann}, {Brownstein},
  {Bundy}, {Dawson}, {Eftekharzadeh}, {Fernandez-Trincado}, {Gao},
  {Hutchinson}, {Jia}, {Jiang}, {Oravetz}, {Pan}, {Paris}, {Ponder}, {Peters},
  {Rogerson}, {Simmons}, {Smith}, \& {Wang}}]{Grier2017}
{Grier}, C.~J., {Trump}, J.~R., {Shen}, Y., {et~al.} 2017, \apj, 851, 21

\bibitem[{{Grier} {et~al.}(2019){Grier}, {Shen}, {Horne}, {Brandt}, {Trump},
  {Hall}, {Kinemuchi}, {Starkey}, {Schneider}, {Ho}, {Homayouni}, {I-Hsiu Li},
  {McGreer}, {Peterson}, {Bizyaev}, {Chen}, {Dawson}, {Eftekharzadeh}, {Guo},
  {Jia}, {Jiang}, {Kneib}, {Li}, {Li}, {Nie}, {Oravetz}, {Oravetz}, {Pan},
  {Petitjean}, {Ponder}, {Rogerson}, {Vivek}, {Zhang}, \& {Zou}}]{Grier2019}
{Grier}, C.~J., {Shen}, Y., {Horne}, K., {et~al.} 2019, \apj, 887, 38

\bibitem[{{Homayouni} {et~al.}(2019){Homayouni}, {Trump}, {Grier}, {Shen},
  {Starkey}, {Brandt}, {Fonseca Alvarez}, {Hall}, {Horne}, {Kinemuchi}, {I-Hsiu
  Li}, {McGreer}, {Sun}, {Ho}, \& {Schneider}}]{Homayouni2019}
{Homayouni}, Y., {Trump}, J.~R., {Grier}, C.~J., {et~al.} 2019, \apj, 880, 126

\bibitem[{{Homayouni} {et~al.}(2020){Homayouni}, {Trump}, {Grier}, {Horne},
  {Shen}, {Brandt}, {Dawson}, {Alvarez}, {Green}, {Hall}, {Hern{\'a}ndez
  Santisteban}, {Ho}, {Kinemuchi}, {Kochanek}, {Li}, {Peterson}, {Schneider},
  {Starkey}, {Bizyaev}, {Pan}, {Oravetz}, \& {Simmons}}]{Homayouni2020}
---. 2020, \apj, 901, 55

\bibitem[{{Jiang} {et~al.}(2017){Jiang}, {Green}, {Greene}, {Morganson},
  {Shen}, {Pancoast}, {MacLeod}, {Anderson}, {Brandt}, {Grier}, {Rix}, {Ruan},
  {Protopapas}, {Scott}, {Burgett}, {Hodapp}, {Huber}, {Kaiser}, {Kudritzki},
  {Magnier}, {Metcalfe}, {Tonry}, {Wainscoat}, \& {Waters}}]{JiangGreen2017}
{Jiang}, Y.-F., {Green}, P.~J., {Greene}, J.~E., {et~al.} 2017, \apj, 836, 186

\bibitem[{{Kammoun} {et~al.}(2021){Kammoun}, {Papadakis}, \&
  {Dov{\v{c}}iak}}]{Kammoun2021}
{Kammoun}, E.~S., {Papadakis}, I.~E., \& {Dov{\v{c}}iak}, M. 2021, \mnras,
  arXiv:2103.04892

\bibitem[{{Kelly} {et~al.}(2009){Kelly}, {Bechtold}, \&
  {Siemiginowska}}]{Kelly2009}
{Kelly}, B.~C., {Bechtold}, J., \& {Siemiginowska}, A. 2009, \apj, 698, 895

\bibitem[{{Korista} \& {Goad}(2001)}]{Korista2001}
{Korista}, K.~T., \& {Goad}, M.~R. 2001, \apj, 553, 695

\bibitem[{{Korista} \& {Goad}(2019)}]{Korista2019}
---. 2019, \mnras, 489, 5284

\bibitem[{{Koz{\l}owski}(2016)}]{Kozlowski2016}
{Koz{\l}owski}, S. 2016, \apj, 826, 118

\bibitem[{{Kriss} {et~al.}(2000){Kriss}, {Peterson}, {Crenshaw}, \&
  {Zheng}}]{Kriss2000}
{Kriss}, G.~A., {Peterson}, B.~M., {Crenshaw}, D.~M., \& {Zheng}, W. 2000,
  \apj, 535, 58

\bibitem[{{Krolik} {et~al.}(1991){Krolik}, {Horne}, {Kallman}, {Malkan},
  {Edelson}, \& {Kriss}}]{Krolik1991}
{Krolik}, J.~H., {Horne}, K., {Kallman}, T.~R., {et~al.} 1991, \apj, 371, 541

\bibitem[{{Kuehn} {et~al.}(2008){Kuehn}, {Baldwin}, {Peterson}, \&
  {Korista}}]{Kuehn2008}
{Kuehn}, C.~A., {Baldwin}, J.~A., {Peterson}, B.~M., \& {Korista}, K.~T. 2008,
  \apj, 673, 69

\bibitem[{{Lawther} {et~al.}(2018){Lawther}, {Goad}, {Korista}, {Ulrich}, \&
  {Vestergaard}}]{Lawther2018}
{Lawther}, D., {Goad}, M.~R., {Korista}, K.~T., {Ulrich}, O., \& {Vestergaard},
  M. 2018, \mnras, 481, 533

\bibitem[{{Li} {et~al.}(2019){Li}, {Shen}, {Brandt}, {Grier}, {Hall}, {Ho},
  {Homayouni}, {Horne}, {Schneider}, {Trump}, \& {Starkey}}]{Li2019}
{Li}, J., {Shen}, Y., {Brandt}, W.~N., {et~al.} 2019, \apj, 884, 119

\bibitem[{{Li} {et~al.}(2021){Li}, {Sun}, {Xu}, {Brandt}, {Trump}, {Yu},
  {Wang}, {Xue}, {Cai}, {Gu}, {Homayouni}, {Liu}, {Wang}, {Zhang}, \&
  {Li}}]{Li2021}
{Li}, T., {Sun}, M., {Xu}, X., {et~al.} 2021, arXiv e-prints, arXiv:2104.12327

\bibitem[{{MacLeod} {et~al.}(2010){MacLeod}, {Ivezi{\'c}}, {Kochanek},
  {Koz{\l}owski}, {Kelly}, {Bullock}, {Kimball}, {Sesar}, {Westman}, {Brooks},
  {Gibson}, {Becker}, \& {de Vries}}]{MacLeod2010}
{MacLeod}, C.~L., {Ivezi{\'c}}, {\v{Z}}., {Kochanek}, C.~S., {et~al.} 2010,
  \apj, 721, 1014

\bibitem[{{MacLeod} {et~al.}(2012){MacLeod}, {Ivezi{\'c}}, {Sesar}, {de Vries},
  {Kochanek}, {Kelly}, {Becker}, {Lupton}, {Hall}, {Richards}, {Anderson}, \&
  {Schneider}}]{MacLeod2012}
{MacLeod}, C.~L., {Ivezi{\'c}}, {\v{Z}}., {Sesar}, B., {et~al.} 2012, \apj,
  753, 106

\bibitem[{{Maronna} {et~al.}(2006){Maronna}, {Martin}, \&
  {Yohai}}]{Maronna2006}
{Maronna}, R.~A., {Martin}, R.~D., \& {Yohai}, V.~J. 2006, {Robust Statistics
  (John Wiley \& Sons Ltd)}

\bibitem[{{McHardy} {et~al.}(2014){McHardy}, {Cameron}, {Dwelly}, {Connolly},
  {Lira}, {Emmanoulopoulos}, {Gelbord}, {Breedt}, {Arevalo}, \&
  {Uttley}}]{McHardy2014}
{McHardy}, I.~M., {Cameron}, D.~T., {Dwelly}, T., {et~al.} 2014, \mnras, 444,
  1469

\bibitem[{{McHardy} {et~al.}(2018){McHardy}, {Connolly}, {Horne}, {Cackett},
  {Gelbord}, {Peterson}, {Pahari}, {Gehrels}, {Goad}, {Lira}, {Arevalo},
  {Baldi}, {Brandt}, {Breedt}, {Chand}, {Dewangan}, {Done}, {Elvis},
  {Emmanoulopoulos}, {Fausnaugh}, {Kaspi}, {Kochanek}, {Korista}, {Papadakis},
  {Rao}, {Uttley}, {Vestergaard}, \& {Ward}}]{McHardy2018}
{McHardy}, I.~M., {Connolly}, S.~D., {Horne}, K., {et~al.} 2018, \mnras, 480,
  2881

\bibitem[{{Momcheva} {et~al.}(2017){Momcheva}, {van Dokkum}, {van der Wel},
  {Brammer}, {MacKenty}, {Nelson}, {Leja}, {Muzzin}, \& {Franx}}]{Momcheva2017}
{Momcheva}, I.~G., {van Dokkum}, P.~G., {van der Wel}, A., {et~al.} 2017,
  \pasp, 129, 015004

\bibitem[{{Morgan} {et~al.}(2018){Morgan}, {Hyer}, {Bonvin}, {Mosquera},
  {Cornachione}, {Courbin}, {Kochanek}, \& {Falco}}]{Morgan2018}
{Morgan}, C.~W., {Hyer}, G.~E., {Bonvin}, V., {et~al.} 2018, \apj, 869, 106

\bibitem[{{Mowla} {et~al.}(2019){Mowla}, {van Dokkum}, {Brammer}, {Momcheva},
  {van der Wel}, {Whitaker}, {Nelson}, {Bezanson}, {Muzzin}, {Franx},
  {MacKenty}, {Leja}, {Kriek}, \& {Marchesini}}]{Mowla2019}
{Mowla}, L.~A., {van Dokkum}, P., {Brammer}, G.~B., {et~al.} 2019, \apj, 880,
  57

\bibitem[{{Mudd} {et~al.}(2018){Mudd}, {Martini}, {Zu}, {Kochanek}, {Peterson},
  {Kessler}, {Davis}, {Hoormann}, {King}, {Lidman}, {Sommer}, {Tucker},
  {Asorey}, {Hinton}, {Glazebrook}, {Kuehn}, {Lewis}, {Macaulay}, {Moeller},
  {O'Neill}, {Zhang}, {Abbott}, {Abdalla}, {Allam}, {Banerji},
  {Benoit-L{\'e}vy}, {Bertin}, {Brooks}, {Carnero Rosell}, {Carollo}, {Carrasco
  Kind}, {Carretero}, {Cunha}, {D'Andrea}, {da Costa}, {Davis}, {Desai},
  {Doel}, {Fosalba}, {Garc{\'\i}a-Bellido}, {Gaztanaga}, {Gerdes}, {Gruen},
  {Gruendl}, {Gschwend}, {Gutierrez}, {Hartley}, {Honscheid}, {James},
  {Kuhlmann}, {Kuropatkin}, {Lima}, {Maia}, {Marshall}, {McMahon}, {Menanteau},
  {Miquel}, {Plazas}, {Romer}, {Sanchez}, {Schindler}, {Schubnell}, {Smith},
  {Smith}, {Soares-Santos}, {Sobreira}, {Suchyta}, {Swanson}, {Tarle},
  {Thomas}, {Tucker}, {Walker}, \& {DES Collaboration}}]{Mudd2018}
{Mudd}, D., {Martini}, P., {Zu}, Y., {et~al.} 2018, \apj, 862, 123

\bibitem[{{Mushotzky} {et~al.}(2011){Mushotzky}, {Edelson}, {Baumgartner}, \&
  {Gandhi}}]{Mushotzky2011}
{Mushotzky}, R.~F., {Edelson}, R., {Baumgartner}, W., \& {Gandhi}, P. 2011,
  \apjl, 743, L12

\bibitem[{{Pahari} {et~al.}(2020){Pahari}, {McHardy}, {Vincentelli}, {Cackett},
  {Peterson}, {Goad}, {G{\"u}ltekin}, \& {Horne}}]{Pahari2020}
{Pahari}, M., {McHardy}, I.~M., {Vincentelli}, F., {et~al.} 2020, \mnras, 494,
  4057

\bibitem[{{Peterson}(1993)}]{Peterson1993}
{Peterson}, B.~M. 1993, Publications of the Astronomical Society of the
  Pacific, 105, 247

\bibitem[{{Peterson} {et~al.}(2004){Peterson}, {Ferrarese}, {Gilbert}, {Kaspi},
  {Malkan}, {Maoz}, {Merritt}, {Netzer}, {Onken}, {Pogge}, {Vestergaard}, \&
  {Wandel}}]{Peterson2004}
{Peterson}, B.~M., {Ferrarese}, L., {Gilbert}, K.~M., {et~al.} 2004, \apj, 613,
  682

\bibitem[{{Read} {et~al.}(2020){Read}, {Smith}, {Jarvis}, \&
  {G{\"u}rkan}}]{Read2020}
{Read}, S.~C., {Smith}, D.~J.~B., {Jarvis}, M.~J., \& {G{\"u}rkan}, G. 2020,
  \mnras, 492, 3940

\bibitem[{{Reynolds} \& {Nowak}(2003)}]{Reynolds2003}
{Reynolds}, C.~S., \& {Nowak}, M.~A. 2003, \physrep, 377, 389

\bibitem[{{Richards} {et~al.}(2006){Richards}, {Lacy}, {Storrie-Lombardi},
  {Hall}, {Gallagher}, {Hines}, {Fan}, {Papovich}, {Vanden Berk}, {Trammell},
  {Schneider}, {Vestergaard}, {York}, {Jester}, {Anderson}, {Budav{\'a}ri}, \&
  {Szalay}}]{Richards2006}
{Richards}, G.~T., {Lacy}, M., {Storrie-Lombardi}, L.~J., {et~al.} 2006, \apjs,
  166, 470

\bibitem[{{Runnoe} {et~al.}(2012){Runnoe}, {Brotherton}, \&
  {Shang}}]{Runnoe2012}
{Runnoe}, J.~C., {Brotherton}, M.~S., \& {Shang}, Z. 2012, \mnras, 426, 2677

\bibitem[{{Salvatier} {et~al.}(2016){Salvatier}, {Wiecki{\^a}}, \&
  {Fonnesbeck}}]{Salvatier2016}
{Salvatier}, J., {Wiecki{\^a}}, T.~V., \& {Fonnesbeck}, C. 2016, {PyMC3:
  Python probabilistic programming framework}, , , ascl:1610.016

\bibitem[{{Sergeev} {et~al.}(2005){Sergeev}, {Doroshenko}, {Golubinskiy},
  {Merkulova}, \& {Sergeeva}}]{Sergeev2005}
{Sergeev}, S.~G., {Doroshenko}, V.~T., {Golubinskiy}, Y.~V., {Merkulova},
  N.~I., \& {Sergeeva}, E.~A. 2005, \apj, 622, 129

\bibitem[{{Shakura} \& {Sunyaev}(1973)}]{SS1973}
{Shakura}, N.~I., \& {Sunyaev}, R.~A. 1973, \aap, 500, 33

\bibitem[{{Shappee} {et~al.}(2014){Shappee}, {Prieto}, {Grupe}, {Kochanek},
  {Stanek}, {De Rosa}, {Mathur}, {Zu}, {Peterson}, {Pogge}, {Komossa}, {Im},
  {Jencson}, {Holoien}, {Basu}, {Beacom}, {Szczygie{\l}}, {Brimacombe},
  {Adams}, {Campillay}, {Choi}, {Contreras}, {Dietrich}, {Dubberley},
  {Elphick}, {Foale}, {Giustini}, {Gonzalez}, {Hawkins}, {Howell}, {Hsiao},
  {Koss}, {Leighly}, {Morrell}, {Mudd}, {Mullins}, {Nugent}, {Parrent},
  {Phillips}, {Pojmanski}, {Rosing}, {Ross}, {Sand}, {Terndrup}, {Valenti},
  {Walker}, \& {Yoon}}]{Shappee2014}
{Shappee}, B.~J., {Prieto}, J.~L., {Grupe}, D., {et~al.} 2014, \apj, 788, 48

\bibitem[{{Shen} {et~al.}(2015){Shen}, {Brandt}, {Dawson}, {Hall}, {McGreer},
  {Anderson}, {Chen}, {Denney}, {Eftekharzadeh}, {Fan}, {Gao}, {Green},
  {Greene}, {Ho}, {Horne}, {Jiang}, {Kelly}, {Kinemuchi}, {Kochanek},
  {P{\^a}ris}, {Peters}, {Peterson}, {Petitjean}, {Ponder}, {Richards},
  {Schneider}, {Seth}, {Smith}, {Strauss}, {Tao}, {Trump}, {Wood-Vasey}, {Zu},
  {Eisenstein}, {Pan}, {Bizyaev}, {Malanushenko}, {Malanushenko}, \&
  {Oravetz}}]{Shen2015a}
{Shen}, Y., {Brandt}, W.~N., {Dawson}, K.~S., {et~al.} 2015, The Astrophysical
  Journal Supplement Series, 216, 4

\bibitem[{{Shen} {et~al.}(2016){Shen}, {Horne}, {Grier}, {Peterson}, {Denney},
  {Trump}, {Sun}, {Brandt}, {Kochanek}, \& {Dawson}}]{Shen2016a}
{Shen}, Y., {Horne}, K., {Grier}, C.~J., {et~al.} 2016, \apj, 818, 30

\bibitem[{{Shen} {et~al.}(2019){Shen}, {Hall}, {Horne}, {Zhu}, {McGreer},
  {Simm}, {Trump}, {Kinemuchi}, {Brandt}, {Green}, {Grier}, {Guo}, {Ho},
  {Homayouni}, {Jiang}, {I-Hsiu Li}, {Morganson}, {Petitjean}, {Richards},
  {Schneider}, {Starkey}, {Wang}, {Chambers}, {Kaiser}, {Kudritzki}, {Magnier},
  \& {Waters}}]{Shen2019a}
{Shen}, Y., {Hall}, P.~B., {Horne}, K., {et~al.} 2019, \apjs, 241, 34

\bibitem[{{Soltan}(1982)}]{Soltan1982}
{Soltan}, A. 1982, \mnras, 200, 115

\bibitem[{{Starkey} {et~al.}(2017){Starkey}, {Horne}, {Fausnaugh}, {Peterson},
  {Bentz}, {Kochanek}, {Denney}, {Edelson}, {Goad}, {De Rosa}, {Anderson},
  {Ar{\'e}valo}, {Barth}, {Bazhaw}, {Borman}, {Boroson}, {Bottorff}, {Brandt},
  {Breeveld}, {Cackett}, {Carini}, {Croxall}, {Crenshaw}, {Dalla Bont{\`a}},
  {De Lorenzo-C{\'a}ceres}, {Dietrich}, {Efimova}, {Ely}, {Evans},
  {Filippenko}, {Flatland}, {Gehrels}, {Geier}, {Gelbord}, {Gonzalez},
  {Gorjian}, {Grier}, {Grupe}, {Hall}, {Hicks}, {Horenstein}, {Hutchison},
  {Im}, {Jensen}, {Joner}, {Jones}, {Kaastra}, {Kaspi}, {Kelly}, {Kennea},
  {Kim}, {Kim}, {Klimanov}, {Korista}, {Kriss}, {Lee}, {Leonard}, {Lira},
  {MacInnis}, {Manne-Nicholas}, {Mathur}, {McHardy}, {Montouri}, {Musso},
  {Nazarov}, {Norris}, {Nousek}, {Okhmat}, {Pancoast}, {Parks}, {Pei}, {Pogge},
  {Pott}, {Rafter}, {Rix}, {Saylor}, {Schimoia}, {Schn{\"u}lle}, {Sergeev},
  {Siegel}, {Spencer}, {Sung}, {Teems}, {Turner}, {Uttley}, {Vestergaard},
  {Villforth}, {Weiss}, {Woo}, {Yan}, {Young}, {Zheng}, \& {Zu}}]{Starkey2017}
{Starkey}, D., {Horne}, K., {Fausnaugh}, M.~M., {et~al.} 2017, \apj, 835, 65

\bibitem[{{Starkey} {et~al.}(2016){Starkey}, {Horne}, \&
  {Villforth}}]{Starkey2016}
{Starkey}, D.~A., {Horne}, K., \& {Villforth}, C. 2016, \mnras, 456, 1960

\bibitem[{{Sun} {et~al.}(2015){Sun}, {Trump}, {Shen}, {Brand t}, {Dawson},
  {Denney}, {Hall}, {Ho}, {Horne}, {Jiang}, {Richards}, {Schneider}, {Bizyaev},
  {Kinemuchi}, {Oravetz}, {Pan}, \& {Simmons}}]{Sun2015}
{Sun}, M., {Trump}, J.~R., {Shen}, Y., {et~al.} 2015, \apj, 811, 42

\bibitem[{{Sun} {et~al.}(2020){Sun}, {Xue}, {Brandt}, {Gu}, {Trump}, {Cai},
  {He}, {Lin}, {Liu}, \& {Wang}}]{Sun2020}
{Sun}, M., {Xue}, Y., {Brandt}, W.~N., {et~al.} 2020, \apj, 891, 178

\bibitem[{{van Dokkum}(2001)}]{vanDokkum2001}
{van Dokkum}, P.~G. 2001, \pasp, 113, 1420

\bibitem[{{Vanden Berk} {et~al.}(2001){Vanden Berk}, {Richards}, {Bauer},
  {Strauss}, {Schneider}, {Heckman}, {York}, {Hall}, {Fan}, {Knapp},
  {Anderson}, {Annis}, {Bahcall}, {Bernardi}, {Briggs}, {Brinkmann}, {Brunner},
  {Burles}, {Carey}, {Castander}, {Connolly}, {Crocker}, {Csabai}, {Doi},
  {Finkbeiner}, {Friedman}, {Frieman}, {Fukugita}, {Gunn}, {Hennessy},
  {Ivezi{\'c}}, {Kent}, {Kunszt}, {Lamb}, {Leger}, {Long}, {Loveday}, {Lupton},
  {Meiksin}, {Merelli}, {Munn}, {Newberg}, {Newcomb}, {Nichol}, {Owen}, {Pier},
  {Pope}, {Rockosi}, {Schlegel}, {Siegmund}, {Smee}, {Snir}, {Stoughton},
  {Stubbs}, {SubbaRao}, {Szalay}, {Szokoly}, {Tremonti}, {Uomoto}, {Waddell},
  {Yanny}, \& {Zheng}}]{VandenBerk2001}
{Vanden Berk}, D.~E., {Richards}, G.~T., {Bauer}, A., {et~al.} 2001, \aj, 122,
  549

\bibitem[{{Vestergaard} \& {Wilkes}(2001)}]{Vestergaard2001}
{Vestergaard}, M., \& {Wilkes}, B.~J. 2001, \apjs, 134, 1

\bibitem[{{Wanders} {et~al.}(1997){Wanders}, {Peterson}, {Alloin}, {Ayres},
  {Clavel}, {Crenshaw}, {Horne}, {Kriss}, {Krolik}, {Malkan}, {Netzer},
  {O'Brien}, {Reichert}, {Rodr{\'\i}guez-Pascual}, {Wamsteker}, {Alexander},
  {Anderson}, {Benitez}, {Bochkarev}, {Burenkov}, {Cheng}, {Collier},
  {Comastri}, {Dietrich}, {Dultzin-Hacyan}, {Espey}, {Filippenko}, {Gaskell},
  {George}, {Goad}, {Ho}, {Kaspi}, {Kollatschny}, {Korista}, {Laor},
  {MacAlpine}, {Mignoli}, {Morris}, {Nandra}, {Penton}, {Pogge}, {Ptak},
  {Rodr{\'\i}guez-Espinoza}, {Santos-Lle{\'o}}, {Shapovalova}, {Shull},
  {Snedden}, {Sparke}, {Stirpe}, {Sun}, {Turner}, {Ulrich}, {Wang}, {Wei},
  {Welsh}, {Xue}, \& {Zou}}]{Wanders1997}
{Wanders}, I., {Peterson}, B.~M., {Alloin}, D., {et~al.} 1997, The
  Astrophysical Journal Supplement Series, 113, 69

\bibitem[{{Yu} {et~al.}(2020){Yu}, {Martini}, {Davis}, {Gruendl}, {Hoormann},
  {Kochanek}, {Lidman}, {Mudd}, {Peterson}, {Wester}, {Allam}, {Annis},
  {Asorey}, {Avila}, {Banerji}, {Bertin}, {Brooks}, {Buckley-Geer}, {Calcino},
  {Rosell}, {Carollo}, {Kind}, {Carretero}, {Cunha}, {D'Andrea}, {Costa}, {De
  Vicente}, {Desai}, {Diehl}, {Doel}, {Eifler}, {Flaugher}, {Fosalba},
  {Frieman}, {Garc{\'\i}a-Bellido}, {Gaztanaga}, {Glazebrook}, {Gruen},
  {Gschwend}, {Gutierrez}, {Hartley}, {Hinton}, {Hollowood}, {Honscheid},
  {Hoyle}, {James}, {Kim}, {Krause}, {Kuehn}, {Kuropatkin}, {Lewis}, {Lima},
  {Macaulay}, {Maia}, {Marshall}, {Menanteau}, {Miquel}, {M{\"o}ller},
  {Plazas}, {Romer}, {Sanchez}, {Scarpine}, {Schubnell}, {Serrano}, {Smith},
  {Smith}, {Soares-Santos}, {Sobreira}, {Suchyta}, {Swann}, {Swanson}, {Tarle},
  {Tucker}, {Tucker}, \& {Vikram}}]{Yu2020}
{Yu}, Z., {Martini}, P., {Davis}, T.~M., {et~al.} 2020, \apjs, 246, 16

\bibitem[{{Zu} {et~al.}(2011){Zu}, {Kochanek}, \& {Peterson}}]{Zu2011}
{Zu}, Y., {Kochanek}, C.~S., \& {Peterson}, B.~M. 2011, \apj, 735, 80

\end{thebibliography}
\end{document}